\documentclass[prb,aps,reprint]{revtex4-1}

  \usepackage{times}

  \usepackage{amsmath}
  \usepackage{amsfonts}
  \usepackage{amssymb}
  \usepackage{graphicx}
  \usepackage[dvipsnames]{xcolor}

  \usepackage{hyperref}

  \graphicspath{{./}{./Figures/}{./result-Figures/}}
  \newcommand{\bns}{\textit{h}-BN~}
  \newcommand{\bn}{\textit{h}-BN}

\begin{document} 

\title{Benchmarking van der Waals-treated DFT: The case of hexagonal boron nitride and graphene on Ir(111)}

\author{Fabian Schulz}
\altaffiliation{Present address: IBM Research - Zurich Research Laboratory, S\"aumerstrasse 4, CH-8803 R\"uschlikon, Switzerland}
\affiliation{Department of Applied Physics, Aalto University School of
  Science, P.O. Box 15100, FI-00076 Aalto, Finland}

\author{Peter Liljeroth}
\affiliation{Department of Applied Physics, Aalto University School of
  Science, P.O. Box 15100, FI-00076 Aalto, Finland}

\author{Ari P Seitsonen}
\altaffiliation{To whom correspondence should be addressed; E-mail: Ari.P.Seitsonen@iki.fi.}
\affiliation{D\'{e}partement de Chimie, \'{E}cole Normale  Sup\'{e}rieure, 24 rue Lhomond, F-75005 Paris, France}
\affiliation{Universit\'{e} de recherche Paris-Sciences-et-Lettres, Sorbonne Universit\'{e}, Centre National de la Recherche Scientifique}

\begin{abstract}
There is enormous recent interest in weak, van der Waals-type (vdW) interactions due to their fundamental relevance for two-dimensional materials and the so-called vdW heterostructures. Tackling this problem using computer simulation is very challenging due to the non-trivial, non-local nature of these interactions. We benchmark different treatments of London dispersion forces within the density functional theory (DFT) framework on hexagonal boron nitride or graphene monolayers on Ir(111) by comparing the calculated geometries to a comprehensive set of experimental data. The geometry of these systems crucially depends on the interplay between vdW interactions and wave function hybridization, making them excellent test cases for vdW-treated DFT. Our results show strong variations in the calculated atomic geometry. While some of the approximations reproduce the experimental structure, this is rather based on \textit{a posteriori} comparison with the ``target results''. General predictive power in vdW-treated DFT is not achieved yet and might require new approaches.
\end{abstract}

\date{\today}

\maketitle

\section{Introduction}

The tremendous success of density functional theory (DFT) together with the continuously growing computing power have laid out the path for the rapidly expanding field of computational materials science. Rather than being employed only in a complementary manner to support the interpretation of experimental data, DFT is now widely used to screen large numbers of structures and compounds,\cite{Yang2012_NatMat,Bjorkman2013_PRX,Jain2016_NatRevMat,Bernevig2017_Nature,Mounet2018_NatNano} with the goal to guide experimentalists in their search and synthesis of new functional materials. In addition, DFT for computational materials science is increasingly combined with machine learning methods,\cite{Behler2016_JChemPhys, Jacobsen2018_PRL, Deringer2018_PRL, Yamashita2018_PRM} which further increases automation and decreases human oversight. This growing trust in and reliance on DFT calculations suggests wide-spread predictive power. While DFT is an \textit{exact} theory -- within the assumptions when deriving it, such as Born-Oppenheimer approximation -- in realistic calculations, however, an approximation to the exchange-correlation functional has to be made. Thus, at the current state of DFT, such an assumption of general applicability of DFT in practice should not be made \textit{a priori} but checked and validated for each class of materials and their properties separately.

For example, DFT-based materials discovery has been extensively used for two-dimensional materials,\cite{Bjorkman2013_PRX,Mounet2018_NatNano} which have recently sparked intense theoretical and experimental interest. These layered materials can be combined in the so-called van der Waals (vdW) heterostructures,\cite{Geim2013_review,Novoselov2106_science} which enables one to tune their structural and electronic behaviour. Because of the absence of covalent bonds between the layers in such heterostructures, vdW interactions are crucial in determining their properties, in contrast to normal, more isotropic solids.\cite{Rehr_1975_a,Pyykkoe_2004_a,Lin_2009_a,Schmidt_2009_a} For example, band hybridization due to interlayer hopping can lead to emergence of superconductivity or strongly-correlated insulating behaviour in certain vdW stacks.\cite{Cao2018_Nature_a, Cao2018_Nature_b}

Another field of research where the interplay of vdW interactions with wave function hybridization is of paramount importance is the adsorption of low-dimensional objects on metal surfaces. Here, the balance between these two interactions governs whether an adsorbate is physisorbed or chemisorbed, and thus also determines adsorption geometry and interfacial electronic properties.\cite{Ertl,Zangwill}

The importance of the vdW interactions - or more precisely the London dispersion forces - in such systems has been recognised for over a decade. However, their computational modelling with the DFT method remains challenging, because these interactions are purely non-local in nature and much weaker than the interactions between chemically bound atoms. Nevertheless, in recent years there has been considerable progress in the treatment of vdW interactions within DFT (in the following referred to as vdW DFT methods). This ranges from semi-empirical approaches with an explicit interaction term depending on the ionic coordinates and usually no dependency on the electronic structure, to density functionals that contain a non-local dependency of the correlation energy on the density.\cite{Klimes2012_JChemPhys,Berland2015_RepProgPhys,Grimme2016_ChemRev} Briefly, in the former case, the system is forced into a geometry that does not constitute a minimum of the Kohn-Sham total energy expression that arises from only the explicit electronic structure and the nucleus-nucleus repulsion. In the latter, the exact form of the vdW energy is known only in the asymptotic limit, which does not correspond to the realistic case where the tails of the electronic densities of the constituents already start to overlap. Therefore, approximations have to be made, and the final expression is not unique, similar to the collection of generalised gradient approximations (GGAs). Consequently, these approximations and their combinations with different exchange and correlation functionals need to be tested and validated.

Many benchmark studies of vdW DFT methods focus on small molecular systems, where reference data can be obtained from highly accurate quantum chemistry calculations.\cite{Klimes_2010_a, Vydrow2012_JCHemTheoComp, DiLabio2013_PCCP} Assessing vdW DFT methods in larger systems is challenging, because their size is prohibitive for quantum chemistry methods. For purely layered compounds, there has been extensive benchmarking against results from the random phase approximation (RPA).\cite{Bjorkman2012_PRL, Bjorkman2014_JChemPhys} If the test system becomes more complex, \textit{e.g.} by including adsorption phenomena, comparison with RPA calculations is feasible only in cases were the unit cell is small.\cite{Mittendorfer2011_PRB, Zhang2014_JChemPhys} In addition, quantum Monte Carlo simulations have recently emerged as an alternative reference for vdW DFT methods.\cite{Ganesh2014_JChemTheoComp, Shulenburger2015_NanoLett} As the unit cells get larger and the complexity of the systems increases, all the computational reference methods discussed above become too expensive. This is particularly true if chemically different materials are involved. Instead, benchmarking vdW DFT methods in such systems then requires reliable experimental data to compare with. Examples of such studies include the calculation of the adsorption height of benzene on various metal surfaces\cite{Yildirim2013_JPhysChemC} as well as of graphene on Ni(111) and Pt(111),\cite{Hamada2010_PRB} two systems where the graphene layer is essentially flat.\cite{Gamo1997_SurfSci,Sutter2009_PRB} For further previous benchmark studies, see Refs.~\citenum{Klimes2012_JChemPhys, Berland2015_RepProgPhys, Grimme2016_ChemRev} and references therein.

Here, we apply DFT to two experimentally well-characterised systems of two-dimensional layers adsorbed on a metal surface,\cite{Auwaerter2019_SSR, Wintterlin2009_SS} namely monolayer hexagonal boron nitride (\bn) on Ir(111) and graphene (gr) on Ir(111), to benchmark the different approximations to the exchange-correlation and vdW terms. These two systems are very challenging for vdW DFT for two reasons: (i) Due to the lattice mismatch between \bns or gr and the substrate, moir\'{e} patterns are formed where the stacking between overlayer and metal surface varies continuously. This leads to very large unit cells, which makes the calculations computationally very expensive. (ii) In addition, the alternating atomic registry with the substrate results in different adsorption strengths along the moir\'e unit cell and consequently, a corrugation of the \bns and gr overlayers. The final geometry of the moir\'e superstructure then crucially depends on the subtle interplay between vdW interactions and wave function overlap leading to charge transfer and chemical bonding. This second point makes these two systems ideal candidates to compare and benchmark vdW DFT, because the modelling of these effects depends on the chosen exchange-correlation functional and the treatment of the vdW interactions therein. 

We test a large number of different vdW DFT methods, and for some of them, we also investigate the influence of varying the parametres in the model, \textit{e.g.} different starting geometries or number of substrate layers. The calculated geometries are compared to experimental results, in particular the adsorption height and corrugation of the overlayers, which represent very sensitive measures for the predictive power of vdW DFT. For the experimental reference values, we draw upon previously published data as well as a new set of noncontact atomic force microscopy experiments conducted specifically within this study.

It should be noted that some of the methods tested here were parametrised on benchmark calculations for small molecular dimers or restricted classes of systems (where they work well), and thus one must not expect them to perform reliably on our chosen model systems. However, the literature shows that many vdW DFT methods are nevertheless frequently employed for calculations beyond what they were parametrised for (please see \textit{e.g.} Table \ref{tbl:one}). After all, the ultimate goal is to devise an approach that works well for all systems, rather than developing a new method for each class of systems. Thus, it is all the more important to critical assess the performance of existing methods for more complex systems.

\section{Previous studies of \bn/Ir(111) and gr/Ir(111)} Before we can assess the different vdW DFT methods, we need to establish a reference to compare the DFT results with. In the following, we provide a brief overview of experimental studies on the adsorption height and moir\'e corrugation of \bn/Ir(111) and gr/Ir(111), as well as some of the previous DFT results (please see Table~\ref{tbl:one}). 

Determining the geometry of such systems is challenging not only from a computational point of view; accessing the adsorption configuration experimentally in a reliable manner is difficult as well. Typical approaches include sample averaging techniques such as dynamic low-energy electron diffraction [LEED-$I$($V$)], x-ray standing wave measurements (XSW) and surface x-ray diffraction (SXRD) and local probes such as scanning tunneling microscopy (STM) and noncontact atomic force microscopy (nc-AFM). While averaging techniques can be affected by surface imperfections such as wrinkles, surface roughness or impurities, local probe techniques often suffer from weak statistical power. It is thus desirable to obtain experimental data for a given system from both approaches of measurement, which ideally should agree with each other.

Many of the above techniques have been extensively demonstrated and compared with DFT calculations for gr/Ir(111).\cite{Busse2011_PRL, Sampsa2013_PRB, Voloshina2013_SciRep, Runte2014_PRB, Liu2014_NanoLett, zumHagen_2016_a} Here, XSW measurements by Busse \textit{et al.}\cite{Busse2011_PRL} yielded a mean adsorption height of 3.38~\AA, in good agreement with DFT calculations presented in the same publication. For the moir\'e corrugation, XSW yielded lower bounds of 0.4 to 1.0~\AA, depending on the coverage (where the largest value corresponds to a full gr monolayer).\cite{Runte2014_PRB} While the lowest value of 0.4~\AA{} is in agreement with LEED-$I$($V$) and nc-AFM results\cite{Sampsa2013_PRB} (to be discussed below) and many of the DFT data,\cite{Busse2011_PRL, Liu2014_NanoLett, Voloshina2013_SciRep, Kumar2017_ACSNano} the remaining values are at odds with most other results, a discrepancy which was attributed to stress in the graphene layer.\cite{Runte2014_PRB} The analysis of XSW data is also not straight-forward, as it requires assumptions about the height distribution within the gr (or \bn) layer and the quality of the prepared surface. SXRD experiments by Jean \textit{et al.}\cite{Jean2015_PRB} yielded essentially the same mean adsorption height (3.39 \AA) and a corrugation of 0.379 \AA, which agrees with the lowest value found with XSW.

While the local probes allow in principle direct access to the moir\'e corrugation,  STM topography is always a convolution of structural and electronic sample properties [for gr/Ir(111) and \bn/Ir(111), this effect can lead even to an inverted apparent moir\'e corrugation\cite{NDiaye2008_NewJPhys,Schulz_2014_a}]. In contrast to STM, nc-AFM measurements are expected to yield values approaching the topographic corrugation.\cite{Boneschnascher2012_ACSNano, Sampsa2013_PRB} The most sophisticated nc-AFM experiments on gr/Ir(111) were carried out by H\"am\"al\"ainen \textit{et al.},\cite{Sampsa2013_PRB} who used a carbon monoxide-functionalised tip to probe the repulsive force regime above the sample and measured a moir\'e corrugation of 0.47~\AA. In the same publication, LEED-$I$($V$) experiments were reported, which yielded a corrugation of 0.43~\AA{} and a mean adsorption height of the gr layer of 3.39~\AA.\cite{Sampsa2013_PRB} The corrugation is in very good agreement with the nc-AFM value, and there is overall good agreement with many of the DFT results.

In the case of \bn/Ir(111), XSW measurements carried out by Farwick zum Hagen \textit{et al.}\cite{zumHagen_2016_a} yielded a lower bound of the moir\'e corrugation of 1.5~\AA{}, in agreement with DFT calculations presented in the same publication of 1.50~\AA{}. It is also close to the DFT value reported by Liu \textit{et al.}\cite{Liu2014_NanoLett} of 1.40~\AA{}, but significantly larger than the DFT corrugation obtained by Schulz \textit{et al.}\cite{Schulz_2014_a} of 0.34~\AA{}. LEED-$I$($V$) and nc-AFM data for the moir\'e adsorption geometry of \bn/Ir(111) are lacking thus far.


\section{Methods}

\subsection{nc-AFM measurements} Details of the sample preparation and nc-AFM experiments are given the Supplementary Material (SM).\cite{Supplementary_Material} Briefly, monolayer \textit{h}-BN on Ir(111) was grown by low-pressure high-temperature chemical vapour deposition under ultra-high vacuum (UHV) conditions (base pressure 10$^{-10}$~mbar) as described in Ref.~\citenum{Schulz_2014_a}. nc-AFM measurements were carried out in a Createc LT-STM/AFM equipped with qPlus tuning fork sensor\cite{Giessibl1998_APL} housed within the same UHV system at a temperature of 5 K. The attractive short-range interactions between the probe tip and the \textit{h}-BN surface were minimised by passivating the tip apex by a carbon monoxide molecule (CO). \cite{Bartels1997_APL,Gross2009_Science,Schulz2018_ACSNano} 

\subsection{DFT calculations} Details of the DFT calculations are given in the SM. We have used two codes for the DFT calculation, \texttt{CP2k} (\url{https://www.CP2k.org/})\cite{Hutter_2014_a} and \texttt{Quantum ESPRESSO} (QE)\cite{Giannozzi_2009_a} (\url{https://www.Quantum-ESPRESSO.org/}); if not otherwise mentioned, the code used was \texttt{CP2k}. In general, we include the vdW interactions to the total energy either in a semi-empirical manner, \textit{ie} an additional term in the total energy that includes or does not the electron density, or by employing a density functional in the exchange and correlation (XC) term. The different treatments employed are listed in the SM.

\section{Results}

%
%

\subsection{nc-AFM measurements on \bn/Ir(111).}
To add a local probe measurement of the \bn/Ir(111) moir\'e corrugation, we have carried out low-temperature nc-AFM experiments with CO-functionalised tips.\cite{Gross2009_Science} The details of the nc-AFM experiments and the sample preparation can be found in the Supporting Material.

\begin{figure}[h]
	\centering
	\includegraphics [width=\textwidth] {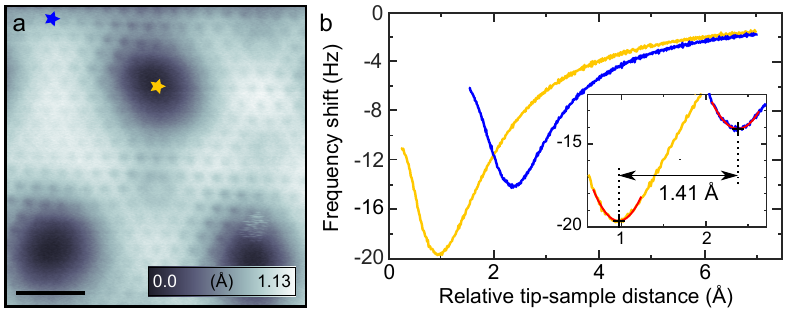}
	\caption{\textbf{Experimental nc-AFM results on \bn/Ir(111).} (\textbf{a}) Constant-$\Delta f$ nc-AFM image of \bn/Ir(111) acquired with a CO-functionalised tip. Setpoint: $-$12.0 Hz. Scale bar is 1~nm. (\textbf{b}) Color-coded $\Delta f(z)$ curves recorded at the positions	marked in panel (a). Inset: Second-order polynomial fits (red) around the $\Delta f$ minima.}
	\label{fig:exp}
\end{figure}

Figure~\ref{fig:exp}a shows an nc-AFM image of \bn/Ir(111) recorded in the constant-$\Delta f$ mode with a CO-passivated tip, yielding the expected moir\'e superstructure composed of depressions arranged in a hexagonal lattice. In addition, the weakly adsorbed regions of the \bns show atomic contrast, revealing the honeycomb lattice of the B and N atoms. There is no atomic contrast in the moir\'e depressions due to long-range vdW interactions between the tip and Ir(111) substrate, which affect the tip-sample distance feedback.\cite{Sun2011_PRB}

Figure~\ref{fig:exp}b shows the measured frequency shift as a function of the tip-sample distance [$\Delta f(z)$] at a moir\'e depression and in the surrounding region. It can be seen that the $\Delta f$ minimum above the depression is not only located at a smaller tip-sample distance but also yields a more negative $\Delta f$ value. In regions where the adsorption height is smaller, the contribution of long-range vdW interaction between tip and iridium substrate is increased, shifting the $\Delta f(z)$ curve to more negative frequency shift values.\cite{Sun2011_PRB} Thus, when imaging on the attractive branch of the $\Delta f(z)$ curve, even in the constant-$\Delta f$ mode with a chemically inert tip apex, the tip-\bns distance will be larger over the moir\'e depressions than over the surrounding regions. Consequently, $\Delta f$ setpoints sufficient to achieve atomic resolution on the latter region might not yield atomic contrast on the former. At very small tip-sample distances, however, the frequency shift is governed by repulsive interactions and it steeply increases as the distance is further reduced; thus the influence of long-range vdW interactions becomes less pronounced. Indeed, it was suggested that the tip-sample distance corresponding to the minimum in $\Delta f(z)$ curves could be used to measure relative differences in adsorption heights.\cite{Schuler2013_PRL}

Following this reasoning, the inset in Figure~\ref{fig:exp}b shows second-order polynomial fits to the minima, from which we extract their vertical difference as 1.41~\AA. Note that this value is significantly larger than the corrugation observed in the constant $\Delta f$ image in Figure~\ref{fig:exp}a. We have carried out analogous measurements with two other CO tips on two other regions of the sample, which yielded differences of 1.70 and 1.85~\AA, respectively. Taking the average of the three values, we get a moir\'e corrugation of (1.65\,$\pm$\,0.23)~\AA. The scatter of the three values and thus the error for the average value seem quite large, however, it should be noted that \bn/Ir(111) in practice is an incommensurate system.\cite{zumHagen_2016_a} In the related system of gr/Ir(111), nc-AFM measurements showed that as a result, even within a single island, the corrugation of individual moir\'e unit cells varies smoothly by approximately $\pm$8\%, with occasional unit cells yielding even up to 20\% larger corrugation.\cite{Sampsa2013_PRB} Overall, our nc-AFM results are in agreement with the value obtained by Farwick zum Hagen \textit{et al.}\/ from XSW (1.55~\AA),\cite{zumHagen_2016_a} as well as with recent DFT results\cite{Liu2014_NanoLett,zumHagen_2016_a,Schulz2018_ACSNano} and thus confirm that \bn/Ir(111) belongs to the class of large-corrugation moir\'e systems.

%
%
\subsection{DFT: \texorpdfstring{$(1\times{}1)$}{}-\bn/Ir(111).} 
\begin{figure*}[ht]
\begin{center}
\includegraphics[width=150mm]{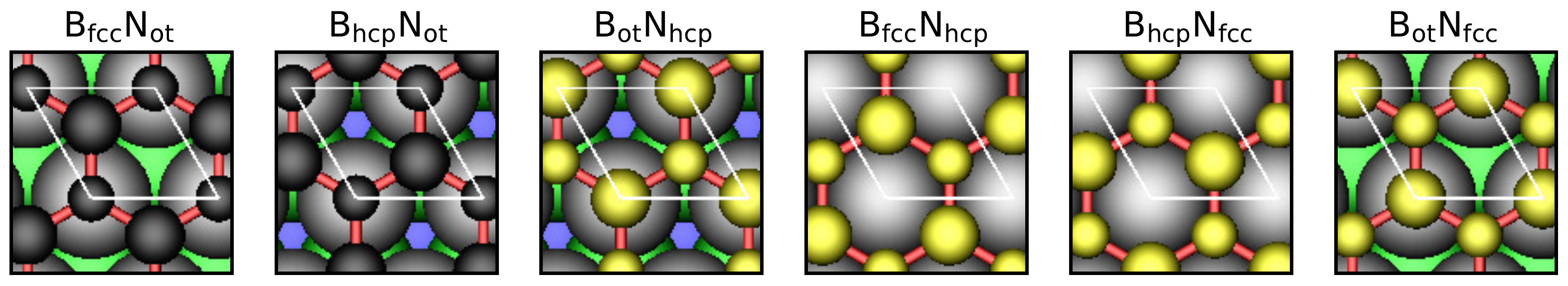}\\
\includegraphics[width=160mm]{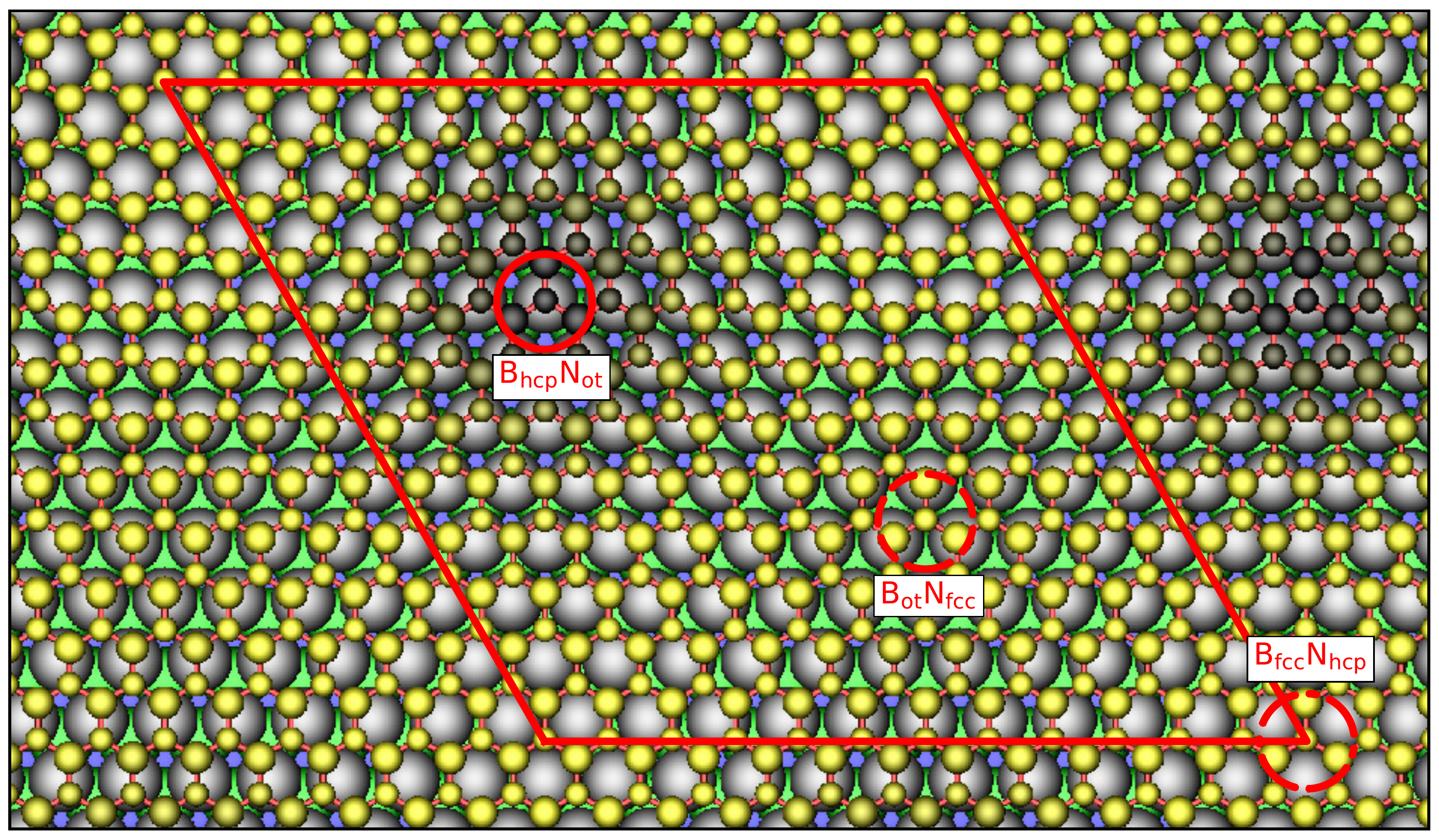}
\includegraphics[width=18mm]{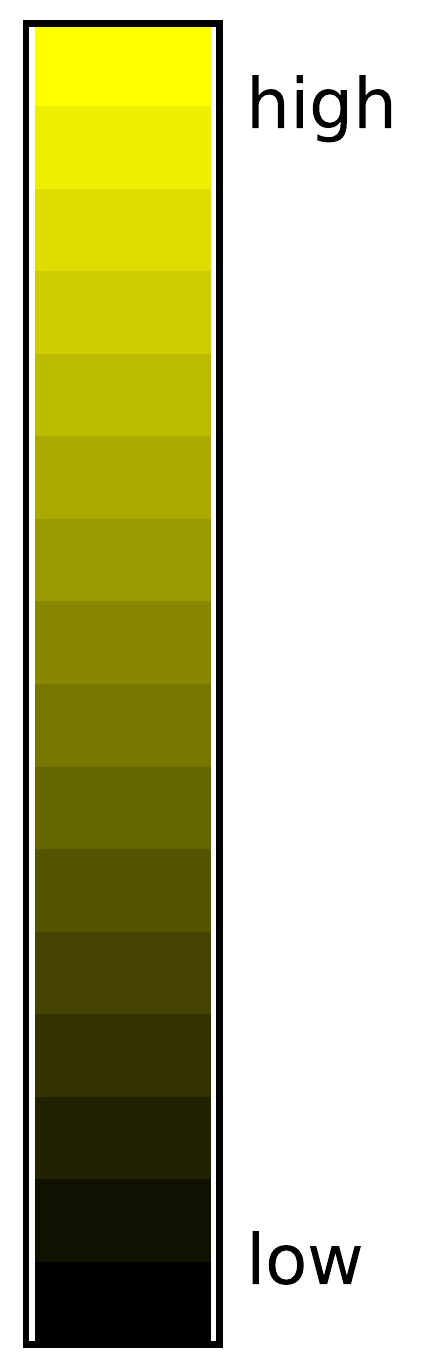}\\
\includegraphics[width=140mm]{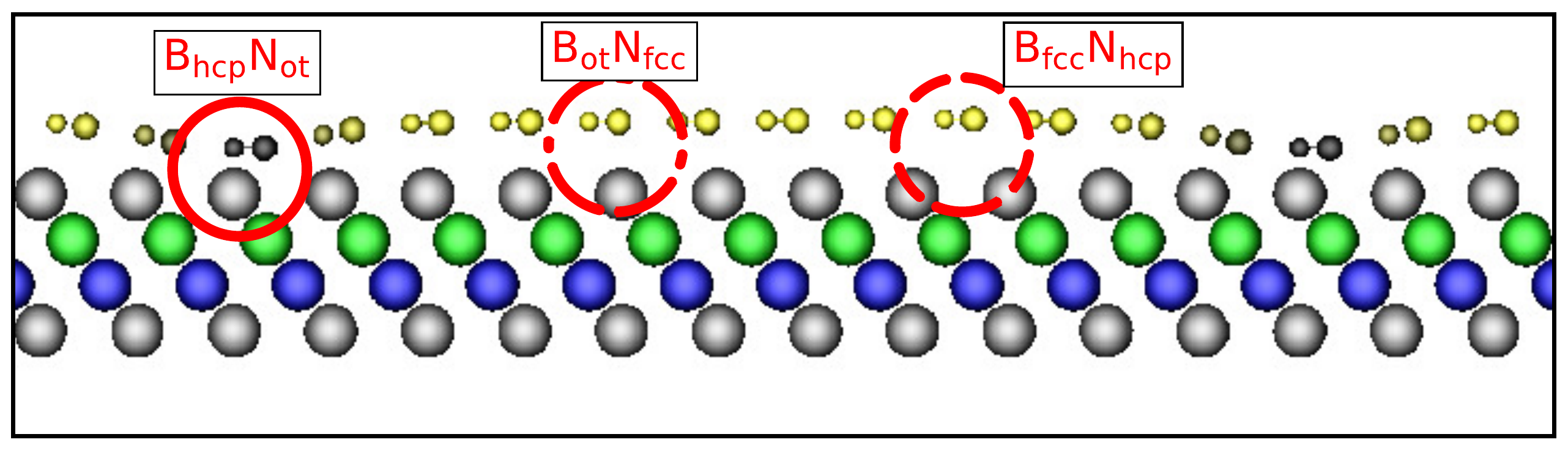}
\caption{\textbf{DFT structures of \bn/Ir(111).} Top panel -- high-symmetry adsorption in the scaled-commensurate structures ($1\times{}1$); middle panel -- top view of the \bn/Ir(111) structure in the 12-on-11 moir\'{e} structure; bottom panel -- the side of the latter structure; lateral adsorption sites are abbreviated as fcc, hcp and ``ot'' = on-top with respect to the fcc(111) termination of the surface. The structures here have been relaxed with the vdW-DF2-rB86 exchange-correlation functional, and the same colour-coding has been used in all panels. B atoms are depicted with large and N with small spheres}
\label{fig:hBN_structure}
\end{center}
\end{figure*}

\begin{figure*}[ht]
\includegraphics[width=\textwidth]{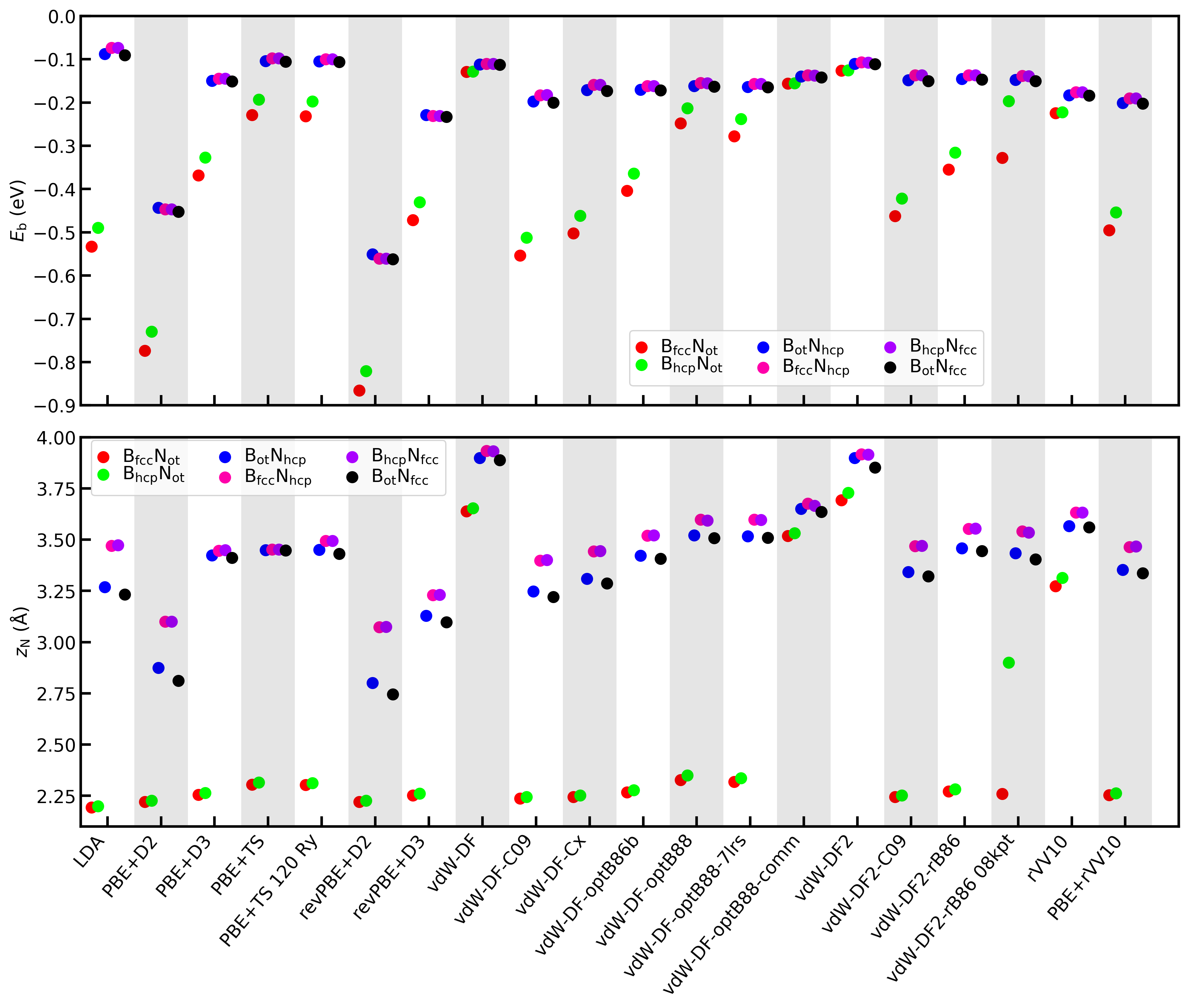}
\caption{\textbf{DFT results on the ($1\times{}1$) commensurate \bn/Ir(111).} Binding energy $E_b$ (top) and height $z_\mathrm{N}$ of N above the top-most substrate layer (bottom) of ($1\times{}1$) commensurate layer, at the experimental lattice constant of Ir(111) from QE-DFT calculations; thus \textit{h}-BN is stretched}
\label{fig:hBN_1x1_Ebind}
\end{figure*}

We now turn to the DFT calculations and how well they can reproduce the experimental geometries. We will devote the major part to \bn/Ir(111), because for this system there is a large spread in the reported DFT results, and thus some controversy. First, we explore the importance of the adsorption site, and the convergence of the results in the moir\'{e} structure by first considering the artificially commensurate ($1\times{}1$)-\bn/Ir(111) structure. 
Despite the strain caused in the \textit{h}-BN layer, this procedure can give information on the differences in the layer height, local buckling and energies at different lateral positions of the overlayer.\cite{GomezDiaz_2013_a} The different structures considered are visualised in Figure~\ref{fig:hBN_structure}. This will naturally be tested later with the calculations of the full moir\'e structure. We also present results obtained with the local density approximation (LDA) even if it does not include vdW interactions or they are not approximated with an additional term, but we want to include the LDA because it is still used in calculations, at least in systems involving graphene. Figure~\ref{fig:hBN_1x1_Ebind} contains the binding energy $E_b$ and the height $z_\mathrm{N}$ of the N atom above the top-most Ir(111) substrate layer. Here the experimental bulk lattice constant $a_{\mathrm{exp}}$ of iridium was used; further details are given in the SM.

We note that the six different domains of high-symmetry positions can be considered as three structures and a rotational centre, so for example B$_\mathrm{fcc}$N$_\mathrm{ot}$ is like B$_\mathrm{hcp}$N$_\mathrm{ot}$ with a rotational domain around the nitrogen atom, differing only in the stacking of the second and third layer of the substrate being inverted.

The immediate observation is that there is not only a large variation in the adsorption strengths of the \textit{h}-BN layer on various lateral adsorption configurations, but also with different treatments of the exchange and correlation. Values of $E_b$ range from almost $-0.1$ to $-0.9$~eV. $z_\mathrm{N}$ varies similarly: Depending on the choice of the XC at the preferred lateral adsorption sites, where N takes the on-top site above the substrate atom, we obtain values from 2.25 to 3.75~\AA. The variations within $E_b$ and $z_\mathrm{N}$ are correlated, so that the approximations to XC that yield the lowest average $E_b$ and/or smallest variations among the different lateral arrangements also place the \textit{h}-BN layer furthest away from the substrate; the weakest adsorption is found with the ``original'' vdW-DF and vdW-DF2 density functionals. These are also the approximations that yield the largest lattice constants (\textit{cf} SM). These two approximations lead in general to too weak binding among the atoms within the system. These shortcomings have been addressed in subsequent approximations by ``fitting'' the lattice constants and later more sophisticated quantities, \textit{e.g.} the adsorption energies of molecules on surfaces.

Considering the variations in $E_b$ and $z_\mathrm{N}$ with a given treatment of XC across the six different high-symmetry adsorption sites, we see that irrespective of the approximation, the placement of N atom on the on-top site yields the strongest binding, with the B$_\mathrm{fcc}$N$_\mathrm{ot}$ slightly preferred over B$_\mathrm{hcp}$N$_\mathrm{ot}$ in energy. The other sites are energetically practically degenerate, even if the B$_\mathrm{ot}$N$_\mathrm{fcc}$ and B$_\mathrm{ot}$N$_\mathrm{hcp}$ tend to reside somewhat closer to the substrate than the arrangements where both atomic species are located at the hollow sites. Thus, the on-top site of either species yields closer adsorption geometry. Independent of the treatment of XC and the lateral adsorption registry, the B atom is closer to the substrate than the N atom.

Given that in the moir\'{e} structures the B and N atoms explore the lateral potential energy surface, not only the high-symmetry sites, in a more or less continuous manner, we can use the trends in $E_b$ and $z_\mathrm{N}$ to ``predict'' the expected trends in the moir\'{e} structures: This suggests small corrugation and large adsorption height with the ``original'' vdW-DF and vdW-DF2 approximations to XC, and the strongest binding and thus the smallest adsorption height with the PBE+D2 and revPBE+D2 approximations; the largest corrugations would then be expected with the PBE+TS, the vdW-DF with C09, Cx, optB86b exchange functionals, and with the PBE+rVV10 approximation.

The results from our two set of calculations with Quantum ESPRESSO (QE) and the PBE+TS treatment (``PBE+TS'', ``PBE+TS 120 Ry'') are independent of the cut-off energy used, indicating a good convergence already at the lower value. Also increasing the number of layers from four to seven, tested with the ``vdW-DF-optB88'' treatment of the XC, leads to the same results. Scaling the lattice constant of the substrate by 11/12 in ``vdW-DF-optB88-comm'' (compared to the equilibrium lattice constant of the bulk), so that the \textit{h}-BN is close to its equilibrium, leads to a very different structure, with almost no preference of the lateral registry. Reducing the number of $k$ points to a grid of 8$\times$8 points leads to small changes in the relative adsorption energies and corrugation.

\subsection{DFT: 12-on-11 \bn/Ir(111) moir\'{e} superstructure.} 
\begin{figure*}[ht]
\includegraphics[width=\textwidth]{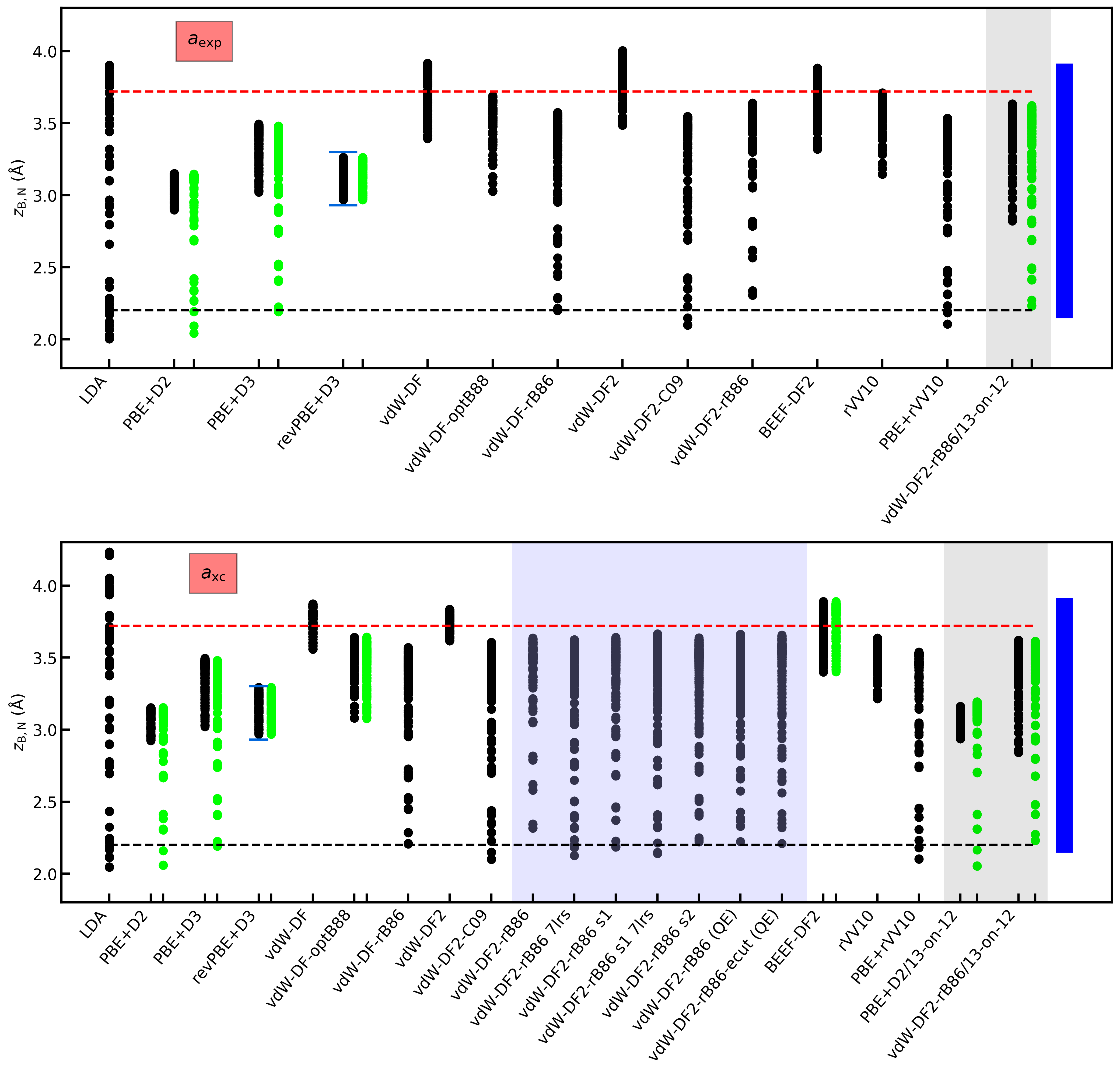}
\caption{\textbf{DFT results on the 12-on-11 and 13-on-12 \bn/Ir(111).} Height of all B and N atoms in 12-on-11 structure above the average height of the top-most layer of the substrate: Top panel, experimental, bottom panel DFT-derived lattice constant; the XSW result by Farwick zum Hagen \textit{et al.} from Ref.~\citenum{zumHagen_2016_a} is marked with the lateral dashed lines, minimum and maximum. Blue horizontal lines mark the minimum and maximum values for a 13-on-12 structure using the revPBE+D3 approximation from Ref.~\citenum{Schulz_2014_a}, and the blue vertical bar on the right indicates the corrugation obtained from the analysis of the AFM data in this publication. With some treatments of the XC we have performed two calculations with different starting geometries -- please see the SM; in this case the second structure is drawn with green markers. The grey-shadowed regions refer to calculations in the 13-on-12 cell, the blue to different calculations with the same approximation.}
\label{fig:hBN_Moire_height}
\end{figure*}

We model the moir\'e structure as commensurate, such that 12$\times$12 cells of \textit{h}-BN match 11$\times$11 cells of Ir(111); we use the notation 12-on-11 in this case. We focus here on the most central results from the DFT calculation, the height of the B and N atoms above the substrate and the corrugation within the \textit{h}-BN layer, given in Figure~\ref{fig:hBN_Moire_height}. With some treatments of the XC, we include two results, originating from the two different procedures of atomic relaxation described in the Methods section. We also include the experimental XSW determination\cite{zumHagen_2016_a} of the minimal and maximal height of the \textit{h}-BN layer with dashed horizontal lines, and the corrugation exctracted from the nc-AFM experiments described above; these act as reference values for the DFT results.

We note that to be more consistent with the experimental XSW analysis, we should use distances from the outer-most layer coordinates of the substrate that have been extrapolated from the bulk coordinates rather than the actual relaxed surface coordinates. We do, however, refer to the latter in trying to minimise the possible confusion of different values, and since the difference is small: For example with the \texttt{CP2k} code and the vdW-DF2-rB86 treatment of the XC approximation, the average distance between the two outer-most substrate layers turns out to be 2.20~\AA, whereas in bulk the distance is 2.22~\AA. Also the magnitude of the corrugation is independent of the reference point of the individual minimum and maximum.

We begin to decipher the results with two general observations that characterize our qualitative DFT results: (i) There is a wide range of heights depending on the different treatments of the XC; the \textit{h}-BN layer is furthest away from the substrate with the same approximations to XC as in the $(1\times{}1)$ structure in Figure~\ref{fig:hBN_1x1_Ebind}, namely vdW-DF and vdW-DF2. In addition, BEEF-DF2 leads to a large average height. (ii) With some treatments of the XC, we find two structures where the forces vanish. Thus there is one stable and one metastable structure. These differ in the magnitude of the corrugation in the \textit{h}-BN layer.

The large variance of the adsorption heights indicates that the approximations to the exchange-correlation term are not reliable \textit{per se}, but a ``calibration'' with good experimental data is necessary. Afterwards, the transferability of the chosen approximation needs to be tested.

That there are two structures which are (meta-)stable in the calculations is an interesting result. This raises the question if both structures are indeed realistic. The approximations that yield the two structures are PBE+D2 and PBE+D3. Some approximations, in particular vdW-DF-rB86, vdW-DF2-C09 and vdW-DF2-rB86 only yield the structure with large corrugation after ionic relaxation. These functionals also result in the best agreement with the experimental corrugation and minimal and maximal heights as measured with nc-AFM and XSW. This suggests that they have the highest ``accuracy'' in the present system and obtaining two different structures is probably unrealistic.

The value of the binding energy per BN unit, defined as
$$
E_b = \left\{ E(\textrm{\textit{h}-BN/Ir}) - \left[ E(\textrm{\textit{h}-BN})
  + E(\textrm{Ir})\right] \right\} / {n_{\mathrm{BN}}}
$$
with the total energies $E$ from the relaxed calculations of the adsorbate
system and the free constituents, is $-$0.164, $-$0.106 and $-$0.145~eV with the DFT+D3, vdW-DF and vdW-DF2-rB86 treatments of the XC effects, respectively. These values are intermediate of the results in the $(1\times{}1)$ cell above. The  values are somewhat smaller than the $-$0.174~eV obtained in recent calculations.\cite{zumHagen_2016_a}

\subsection{DFT: 13-on-12 \bn/Ir(111) moir\'{e} superstructure.} Even if the closest commensurate periodicity of the aligned moir\'{e} pattern has now been established to be 12-on-11 (Refs.~\onlinecite{Schulz_2014_a,zumHagen_2016_a}), we studied also the larger 13-on-12 moir\'{e} pattern to compare with an earlier DFT calculation.\cite{Schulz_2014_a} The results are also illustrated in Figure~\ref{fig:hBN_Moire_height}; the numerical values are also listed in the SM. The resulting geometry is practically the same in both 12-on-11 and 13-on-12 structures with a given treatment of the XC term. Thus, it is clear that the origin of the different corrugations in different DFT calculations is the XC term, and not the moir\'e periodicity as was suggested earlier\cite{zumHagen_2016_a} (with a possible contribution from the different numerical parametres in the calculations).


%
%

\subsection{DFT: 10-on-9 gr/Ir(111) moir\'{e} superstructure.} 
\begin{figure*}[ht]
\includegraphics[width=\textwidth]{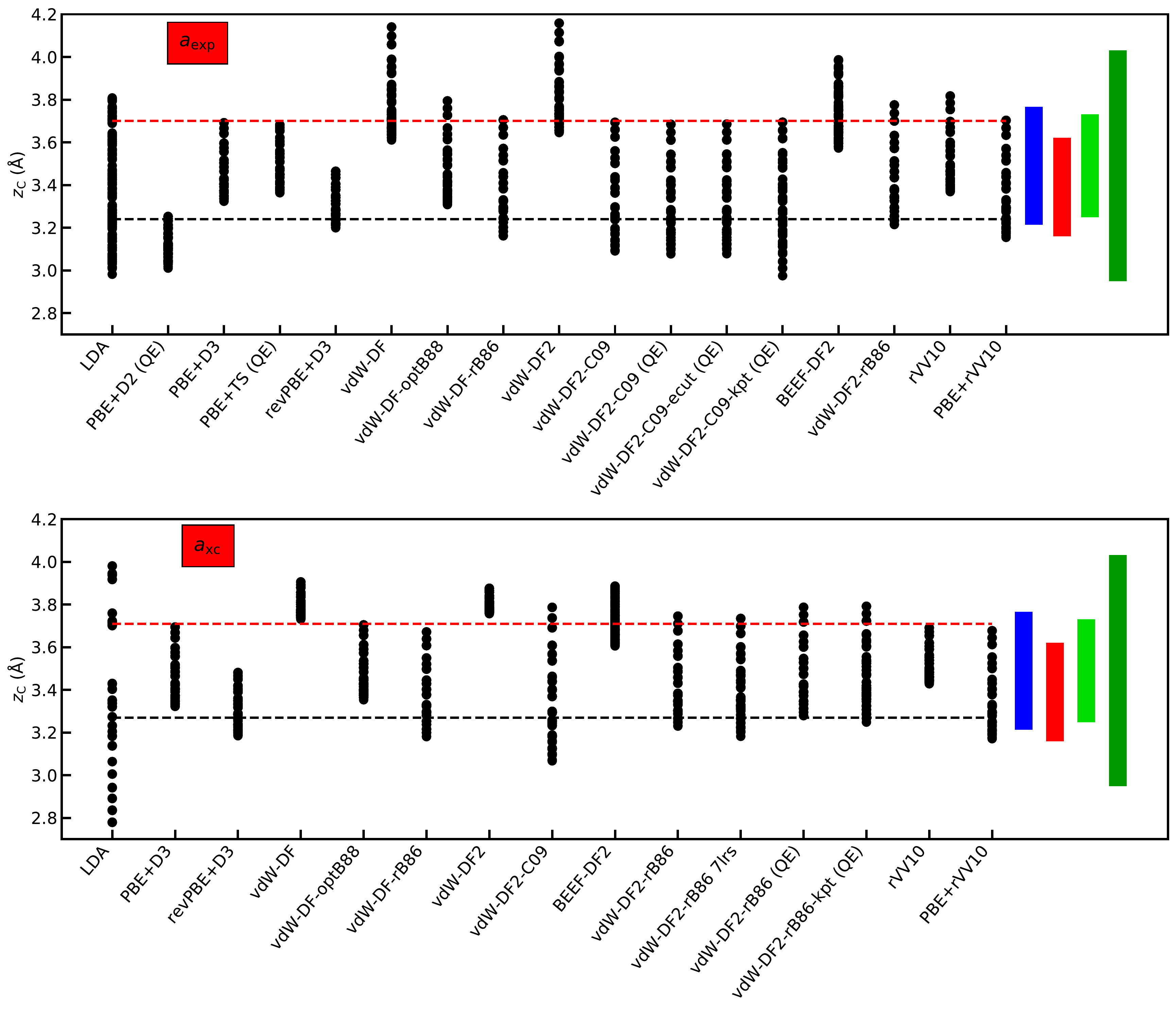}
\caption{\textbf{DFT results on the 10-on-9 gr/Ir(111).} Height of C atoms above the average top-most layer of the substrate in the 10-on-9 structure of gr/Ir(111): Top panel, experimental, bottom panel DFT-derived lattice constant; the LEED-$I$($V$) result by H\"am\"al\"ainen \textit{et al.} from Ref.~\citenum{Sampsa2013_PRB} is marked with the dashed lines, minimum and maximum. The blue bar on the right indicates the corrugation obtained from the analysis of the AFM data in Ref.~\citenum{Sampsa2013_PRB}, the red one from the SXRD data in Ref.~\onlinecite{Jean2015_PRB} and the green ones the extremal values from XSW data in Ref.~\citenum{Runte2014_PRB}.}
\label{fig:gr_Moire_height}
\end{figure*}

We have also checked the structure of graphene on Ir(111) with a variety XC functionals and treatments of vdW interactions. Here, we focus on periodicity 10-on-9 (calculations on $1\times1$ structure can be found in the SM). The height of the carbon atoms above the outer-most layer of Ir(111) are shown in Figure~\ref{fig:gr_Moire_height} together with the results from the experimental derivation of the atomic heights from the LEED-$I$($V$), XSW and nc-AFM experiments.\cite{zumHagen_2016_a,Sampsa2013_PRB} Again several treatments of the XC term have been employed, some at the experimental, some at the equilibrium DFT value of the lattice constant.

Most of the treatments of the exchange-correlation term yield reasonably accurate adsorption heights. The variation in the heights is less pronounced than in the case of \textit{h}-BN/Ir(111), due to the overall smaller corrugation of the graphene layer. Still the same approximations lead to too weak binding, and thus too large adsorption heights (including vdW-DF, vdW-DF2 and BEEF-DF2), therefore binding too little in both type of investigated moir\'{e} structures.

As the lateral unit cell is here smaller than in \textit{h}-BN/Ir(111), the importance of the proper $k$ point sampling in the Brillouin zone is enhanced; also the calculations with more $k$ points are less expensive. We used QE in convergence studies, by increasing the cut-off energy and $k$ point sampling to $2\times{}2$. The former does not lead to any noticeable difference in the height of the carbon atoms, whereas increasing the $k$ point sampling leads to a quantitative difference, where some carbon atoms are adsorbed closer to the surface than with only the $\Gamma$ point.

%
\section{Discussion: Evaluation of the different approaches} With the present results the structure of \textit{h}-BN on Ir(111) becomes clearer, resolving most of the controversies in the past literature: Two different moir\'{e} structures, one with smaller, one with larger corrugation, can be found with some approximations to the exchange and correlation effects in the DFT calculations. The details of the calculations, whether the consistently optimised DFT-XC lattice constant or 12-on-11 or 13-on-12 lateral cell is used, do not affect the results much.

The energy difference $\Delta{}E$ between these two kind of structures indicates how delicate the geometry is: For the semi-empirical methods, sometimes the larger corrugation is preferred -- $a_\mathrm{xc}$-PBE+D2 ($\Delta{}E$ = -0.13~eV), $a_\mathrm{xc}$-PBE+D2/13-on-12 ($\Delta{}E$ = -0.26~eV), sometimes the smaller corrugation $a_\mathrm{xc}$-PBE+D3 ($\Delta{}E$ = +0.13~eV), $a_\mathrm{exp}$-PBE+D3 ($\Delta{}E$ = +0.13~eV), $a_\mathrm{exp}$-PBE+D2 ($\Delta{}E$ = +0.18~eV). Where the vdW-functionals yield two structures, they are energetically almost degenerate, with the $\Delta{}E$ difference being -40~meV or smaller in favour of the larger corrugation. That the $\Delta{}E$ is always so small, at most $\approx$ 0.2~eV in magnitude in such a large moir\'{e} cell, indicates that the energy balance is indeed very sensitive, and we cannot exclude that even slightest numerical issues, or already a finite but low temperature, might turn the balance toward the other kind of corrugation. Thus far, there is no evidence for the experimental observation of the lower corrugation, and it is not clear if it can be realised at all. In the case of \textit{h}-BN on Rh(111), experiments suggest that transition metal adatoms are able to partially weaken the bonding between \textit{h}-BN and metal surface in the moir\'e depressions.\cite{Natterer2012_PRL} This relatively easy vertical modification of the \bns hints at two energetically relatively close-lying structures at different corrugations, and a similar effect could be present in \bn/Ir(111) as well.

Overall, there are in both approaches to model the vdW forces in DFT, semi-empirical and nonlocal, treatments that yield good or very good agreement with the experiment and some that miserably fail. In particular the original vdW-DF and vdW-DF2 yield too weak interaction between the \bns or gr and the substrate, respectively. While this underbinding transfers from one system to the other for certain vdW DFT methods, there is no general trend in the transferability. This is highlighted in the results for vdW-DF-optB88, which show good agreement for gr but completely underestimate the corrugation for \bn. Similarly, while none of the semi-empirical methods performs particularly well for \bn/Ir(111), PBE+D3 yields very good agreement in the case of gr/Ir(111). Within our study, only vdW-DF-rB86, vdW-DF2-rB86 and PBE+rVV10 among the nonlocal treatments yielded good to very good agreement with the experimental results for both gr and \bn. However, that there is overall such a large scatter in the calculated structures is somewhat worrying, and suggests that this is not a universal result.

The corrugations between the adsorption of \bns at the different high-symmetry lattice sites in the ($1\times{}1$)-\bn/Ir(111) unit cell are similar to the range of adsorption heights in the moir\'{e} cell when the \bns has been stretched to be commensurate with the lattice of the substrate, but the results are very different when the \bns is allowed to keep its equilibrium distance and the lattice constant of the metal is changed instead. Therefore, we conclude that the more realistic approach to describe the full moir\'{e} structure is to stretch the \bns than the substrate, as the electronic structure of the latter is considerably changed. On the other hand, this cùould be used in epitaxial, thin metal-on-support over-layers, where the substrate layer has had its in-plane lattice constant changed and thus its electronic structure significantly modified also, to derive new \bns adsorption structures.

In a previous study of gr on Ni(111) and Pt(111) it was found\cite{Hamada2010_PRB} that results obtained with vdW-DF-C09 or vdW-DF2-C09 agree best with the experimental adsorption height, while the original vdW-DF and vdW-DF2 yield considerable underbinding. In our case, vdW-DF2-C09 works well for \bn, but considerably overestimates the corrugation in gr/Ir(111). The original vdW-DF2 also severely underbinds in our study.

Results closer to the experiments are nowadays obtained mainly by ``tuning'' the approximation used in the exchange term, even if the vdW interactions originate purely from the correlation, due to the unpredictable error cancellation between the exchange and correlation. This kind of ``fitting'' to obtain correct structures is an indication of the challenge that the community currently faces while looking for an efficient yet most accurate and ``reliable'' approximation to use. A recent example of comparison of different approaches in solids\cite{Tran_2019_PRM_a} supports our own conclusions. Overall it is thus not sufficient to state that the calculations were performed using \emph{a} treatment of the vdW effects, as this is no guarantee of the accuracy of the results.

%
%
\section{Conclusions}\label{sec:conclusions}

In conclusion, we have performed DFT calculations on \textit{h}-BN and graphene adsorbed on Ir(111) using different approximations to incorporate the vdW interactions. Overall we find a large variation in the corrugation and distance of the \textit{h}-BN layer from the substrate with different treatments of the vdW interactions. This supports the ``common knowledge'' that the choice of the treatment is to some degree more ``ad hoc'' than ``predicted'', and the DFT community is in the process of searching for the ``best'' approach, much like the GGAs have been ``fitted'' over the years -- and like there, the approach to be ``preferred'' can depend on the kind of system under study. Recent attempts try to approach the problem via fulfilling rigorous conditions known for the exact vdW density functional; studies addressing their accuracy in practical cases like the present ones are needed. Future progress requires both quantitative experiments and developments in vdW DFT methodology.

\bibliographystyle{apsrev4-1}
\bibliography{article-corrugation_pure}

\noindent \textbf{Acknowledgements:} \\
\noindent \textbf{Funding:} This research made use of the Aalto Nanomicroscopy Center (Aalto NMC) facilities and was supported by the Academy of Finland (Projects No.~311012 and 314882, and Academy professor funding No.~318995 and the European Research Council (ERC 2017 AdG No.~788185). APS acknowledges the computational resources at CSC, Espoo, project 2000606, and Centro Svizzero di Calcolo Scientifico (CSCS), Lugano, Project uzh11.\\
\noindent \textbf{Author Contributions} The project was initiated by APS, who also carried out all the DFT calculations. FS carried out the nc-AFM experiments. APS and FS wrote the manuscript, with additional input from PL. All authors contributed to the discussion and interpretation of the results.\\
\noindent \textbf{Competing Interests} The authors declare that they have no competing financial interests.\\
\noindent \textbf{Data and materials availability:} Additional data and materials are available online at \url{http://www.apsi.me/Science/Projects/2D-layer/hBN-Ir111/index.html}.

\section*{Figures}

\onecolumngrid

\twocolumngrid

\section*{Tables}

\begin{table}[ht]
\caption{Structural parametres from previous DFT calculations and experiments on \bns and gr on Ir(111).}\vspace{12pt}
\centering
\begin{tabular}{ll|lll}
\hline\hline
& & \multicolumn{1}{c}{corrugation (\AA)} & \multicolumn{1}{c}{distance (\AA)} \\
& Method & $\Delta_\mathrm{BN}$ & $\overline{z}_{h\mathrm{-BN}}$-$\overline{z}_\mathrm{Ir_1}$ & Reference \\
\hline

\bns & vdW-DF2-rB86\,  &  \,1.50  & 3.24           & Ref. \citenum{zumHagen_2016_a}\\
     & revPBE+D3       & \,0.338  & 3.187          & Ref. \citenum{Schulz_2014_a}\\
     & PBE+D2          & \,1.4    & 3.8            & Ref. \citenum{Liu2014_NanoLett}\\
     & XSW             & \,$\geqslant1.5$ & 2.20, 3.72\footnotemark[1] & Ref. \citenum{zumHagen_2016_a}\\
     & nc-AFM          & \,1.65   & N/A            & present work \\
     & & & & \\
     & & $\Delta_\mathrm{CC}$ & $\overline{z}_{\mathrm{C}}$-$\overline{z}_\mathrm{Ir_1}$ &  \\
gr   & vdW-DF          & \,0.35   & 3.41           & Ref. \citenum{Busse2011_PRL}\\
     & vdW-DF2-rB86    & \,0.36   & 3.43           & Ref. \citenum{zumHagen_2016_a}\\
     & PBE+D2          & \,0.2    & 4.2            & Ref. \citenum{Liu2014_NanoLett}\\
     & PBE+TS          & \,0.46   & 3.32           & Ref. \citenum{Kumar2017_ACSNano}\\
     & XSW             & \,$\geqslant(0.4 - 1.0)$\footnotemark[2]  & 3.38   & Refs. \citenum{Runte2014_PRB, Busse2011_PRL} \\
     & SXRD            & \,0.379  & 3.39           & Ref. \citenum{Jean2015_PRB}\\
     & LEED-$I$($V$)   & \,0.43   & 3.39           & Ref. \citenum{Sampsa2013_PRB} \\
     & nc-AFM          & \,0.47   & N/A            & Ref. \citenum{Sampsa2013_PRB} \\
 \hline
 \hline
\end{tabular}
\footnotetext[1]{Average adsorption height of the strongly and weakly interacting species, respectively.}
\footnotetext[2]{Coverage-dependent, where 1.0~\AA{} was found in a full monolayer of gr.}
\label{tbl:one}
\end{table}

\end{document}


\setlength{\baselineskip}{16pt}

\title{Supplementary Material:\texorpdfstring{\\}{}
Benchmarking van der Waals-treated DFT: The case of hexagonal boron nitride and graphene on Ir(111)} 

\author{Fabian Schulz}
\altaffiliation{Present address: IBM Research - Zurich Research Laboratory, S\"aumerstrasse 4, CH-8803 R\"uschlikon, Switzerland}
\affiliation{Department of Applied Physics, Aalto University School of
  Science, P.O. Box 15100, FI-00076 Aalto, Finland}

\author{Peter Liljeroth}
\affiliation{Department of Applied Physics, Aalto University School of
  Science, P.O. Box 15100, FI-00076 Aalto, Finland}

\author{Ari P Seitsonen}
\altaffiliation{To whom correspondence should be addressed; E-mail: Ari.P.Seitsonen@iki.fi.}
\affiliation{D\'{e}partement de Chimie, \'{E}cole Normale  Sup\'{e}rieure, 24 rue Lhomond, F-75005 Paris, France}
\affiliation{Universit\'{e} de recherche Paris-Sciences-et-Lettres, Sorbonne Universit\'{e}, Centre National de la Recherche Scientifique}

\date{\today}

\maketitle

%
%
\section{Methods}\label{sec:methods}

\textbf{nc-AFM measurements.} Monolayer \textit{h}-BN on Ir(111) was grown by low-pressure high-tem\-per\-a\-ture chemical vapour deposition under ultra-high vacuum (UHV) conditions (base pressure 10$^{-10}$~mbar) as described in Ref.~\citenum{Schulz_2014_a}. nc-AFM measurements were carried out in a Createc LT-STM/AFM housed within the same UHV system. The microscope was equipped with a qPlus tuning fork sensor\cite{Giessibl1998_APL} and operated at a temperature of 5~K. The qPlus sensor had a resonance frequency $f_0$ of $\sim$30.68~kHz, a quality factor $Q$ of $\sim$98k and a stiffness $k$ of $\sim$1.8~kN/m. In order to minimise attractive short-range interactions between the probe tip and the \textit{h}-BN surface, the tip apex was passivated by deliberate pick-up of a carbon monoxide molecule (CO) from a Cu(111) surface,\cite{Bartels1997_APL, Gross2009_Science} prior to all measurements. After successful CO pick-up, the Cu(111) sample was exchanged for the \textit{h}-BN/Ir(111) sample.\cite{Boneschnascher2012_ACSNano} All subsequent nc-AFM measurements were acquired in the frequency modulation mode\cite{Albrecht1991_JAP} using an oscillation amplitude of 50~pm and at a sample bias voltage of 0~V.

\textbf{DFT calculations.} We performed total energy calculations using density functional theory (DFT)\cite{Hohenberg_1964_a} within the Kohn-Sham formalism.\cite{Kohn_1965_a} We have used two codes for the DFT calculation, \texttt{CP2k} (\url{http://www.CP2k.org/}) and \texttt{Quantum ESPRESSO} (QE) (\url{http://www.Quantum-ESPRESSO.org/}); the choice of two separate codes gave us the possibility to include more approximations in the total energy functional, and verify the results obtained with the two different kinds of numerical implementations. If not otherwise mentioned, the code used was
\texttt{CP2k}. The details of the calculation are given in the following section. In general, we include the vdW interactions to the total energy either in a semi-empirical manner, \textit{ie} an additional term in the total energy that includes or does not the electron density, or by employing a density functional in the exchange and correlation (XC) term. The XC/vdW treatments employed are listed in Table~\ref{tab:DFT_XC_list}.

\begin{table}[ht]
\centering
\caption{DFT approaches employed in the present work.}\vspace{12pt}
\begin{tabular}{l|llcl}\hline\hline
XC & Exchange & Correlation & vdW & References\\\hline
LDA            & Slater   & PZ     &    & Ref. \citenum{Perdew_1981_a}\\
PBE+D2         & PBE      & PBE    & D2 & Refs. \citenum{Perdew_1996_a,Grimme_2004_a}\\
PBE+D3         & PBE      & PBE    & D3 & Refs. \citenum{Perdew_1996_a,Grimme_2010_a}\\
PBE+TS         & PBE      & PBE    & TS & Refs. \citenum{Perdew_1996_a,Tkatchenko_2009_a}\\
revPBE+D2      & revPBE   & PBE    & D2 & Refs. \citenum{Zhang_1998_a,Perdew_1996_a,Grimme_2004_a}\\
revPBE+D3      & revPBE   & PBE    & D3 & Refs. \citenum{Zhang_1998_a,Perdew_1996_a,Grimme_2010_a}\\
vdW-DF         & revPBE   & LDA+DF1   & & Refs. \citenum{Zhang_1998_a,Dion_2004_a}\\
vdW-DF-C09     & C09      & LDA+DF1   & & Refs. \citenum{Cooper_2010_a,Dion_2004_a}\\
vdW-DF-Cx      & Cx       & LDA+DF1   & & Refs. \citenum{Berland_2014_a,Dion_2004_a}\\
vdW-DF-optB86b & optB86b  & LDA+DF1   & & Refs. \citenum{Klimes_2011_a,Dion_2004_a}\\
vdW-DF-optB88  & optB88   & LDA+DF1   & & Refs. \citenum{Klimes_2010_a,Dion_2004_a}\\
vdW-DF-rB86    & rB86     & LDA+DF1   & & Refs. \citenum{Hamada_2014_a,Dion_2004_a}\\
vdW-DF2        & rPW86    & LDA+DF2   & & Refs. \citenum{Murray_2009_a,Lee_2010_a}\\
vdW-DF2-C09    & C09      & LDA+DF2   & & Refs. \citenum{Cooper_2010_a,Lee_2010_a}\\
vdW-DF2-rB86   & rB86     & LDA+DF2   & & Refs. \citenum{Hamada_2014_a,Lee_2010_a}\\
BEEF-DF2       & BEEF     & scPBE+DF2 & & Refs. \citenum{Wellendorff_2012_a,Perdew_1996_a,Lee_2010_a} \\
rVV10          & rPW86    & PBE+rVV10 & & Refs. \citenum{Perdew_1996_a,Sabatini_2013_a}\\
PBE+rVV10      & PBE      & PBE+rVV10 & & Refs. \citenum{Murray_2009_a,Perdew_1996_a,Sabatini_2013_a}\\
\hline\hline
\end{tabular}
\label{tab:DFT_XC_list}
\end{table}

In addition, we tested different lattice constants and investigated how they influence the final geometry: (i) the experimental lattice constant $a_\mathrm{exp}$ = 3.840~\AA,\cite{Ashcroft_1976_a} (ii) the DFT-optimised value of bulk Ir $a_\mathrm{xc}$ or (iii) the same lattice constant as used in Ref.~\citenum{GomezDiaz_2013_a} of $a_{\mathrm{Ir}} = $~3.801~\AA, obtained by optimising the lattice constant of \textit{h}-BN, $a_{\mathit{h}\mathrm{-BN}} = $~2.688~\AA. We started the optimisation of the structure with $a_\mathrm{xc}$ from a flat layer of \textit{h}-BN or gr at either 2.0 or 2.7~\AA{} above the outer-most layer of the truncated, four-layer slab of Ir(111), and kept the two lowest layers of Ir fixed at the bulk positions; to start the calculation with $a_\mathrm{exp}$ we rescaled isotropically the previously obtained geometry to the experimental lattice constant. In some calculations we initially kept the N atom fixed at the distance 2.0~\AA{}, and when the structure was otherwise converged, we relieved this constraint. Some of these calculations led to a structure with a larger corrugation and a smaller height of the depressions above the substrate than the calculation where also this N atom was allowed to fully relax from the beginning; thus two different geometries were obtained. Some calculations yielded the same geometry independent of the constraint.

In calculations of \textit{h}-BN we mainly used the moir\'{e} periodicity of 12-on-11, but did some calculations also with 13-on-12 so as to be able to compare with the previous results\cite{Schulz_2014_a} and the dependency of the calculated properties on the periodicity.

\textbf{Setup of calculations with \texttt{CP2k}.} The Goedecker-Teter-Hutter (GTH) type of pseudo potentials\cite{Goedecker_1996_a} were employed, with 17 valence electrons in Ir. The basis set to expand the wave functions was DZVP-MOLOPT-SR-GTH. We used a cut-off energy of 700~Ry for expanding the density, with five grids and relative cut-off of 70~Ry. LibXC library\cite{Marques_2012_a} of exchange-correlation functionals was used in some combinations of the exchange and correlation functionals. Only the $\Gamma$ point was included in the reciprocal space, without further sampling of the first Brillouin zone. The Fermi-Dirac broadening of the occupation numbers with ``temperature'' of 300~K was used. Due to the weak interactions and strong requirement for precision, the convergency criterion on the maximum force on any ion was set to 0.0514~meV/\AA.

The determination of the lattice constant was done in a simple cubic cell of 1372 atoms ($7\times{}7\times{}7$ cells \`{a} 4 atoms) with $\Gamma$ point only in the first Brillouin zone.

\textbf{Setup of calculations with \texttt{Quantum ESPRESSO}.} We used the cut-off energy of 30 and 400~Ry to expand the Kohn-Sham orbitals and the augmented electron density, respectively, except in tests of the convergence, where they were 50 and 500~Ry. The projector-augmented wave datasets were obtained from \texttt{pslibrary}.\cite{DalCorso_2014_a} Only the $\Gamma$ point was used in the calculations in the moir\'{e} unit cells, except in the test calculations, where an equi-distant $2\times{}2$ grid was used, centred at the $\Gamma$ point,\cite{Monkhorst_1977_a,Seitsonen_2000_a} together with the Fermi-Dirac broadening of the occupation numbers with a width of 50~meV. Convergence thresholds for the total energy and forces on the ions were chosen as 0.0136~meV and 0.257~meV/\AA, respectively.

%
%
\section{Results on bulk Ir}\label{sec:bulk}

\begin{figure}[h!t]
\centering
\includegraphics[width=130mm]{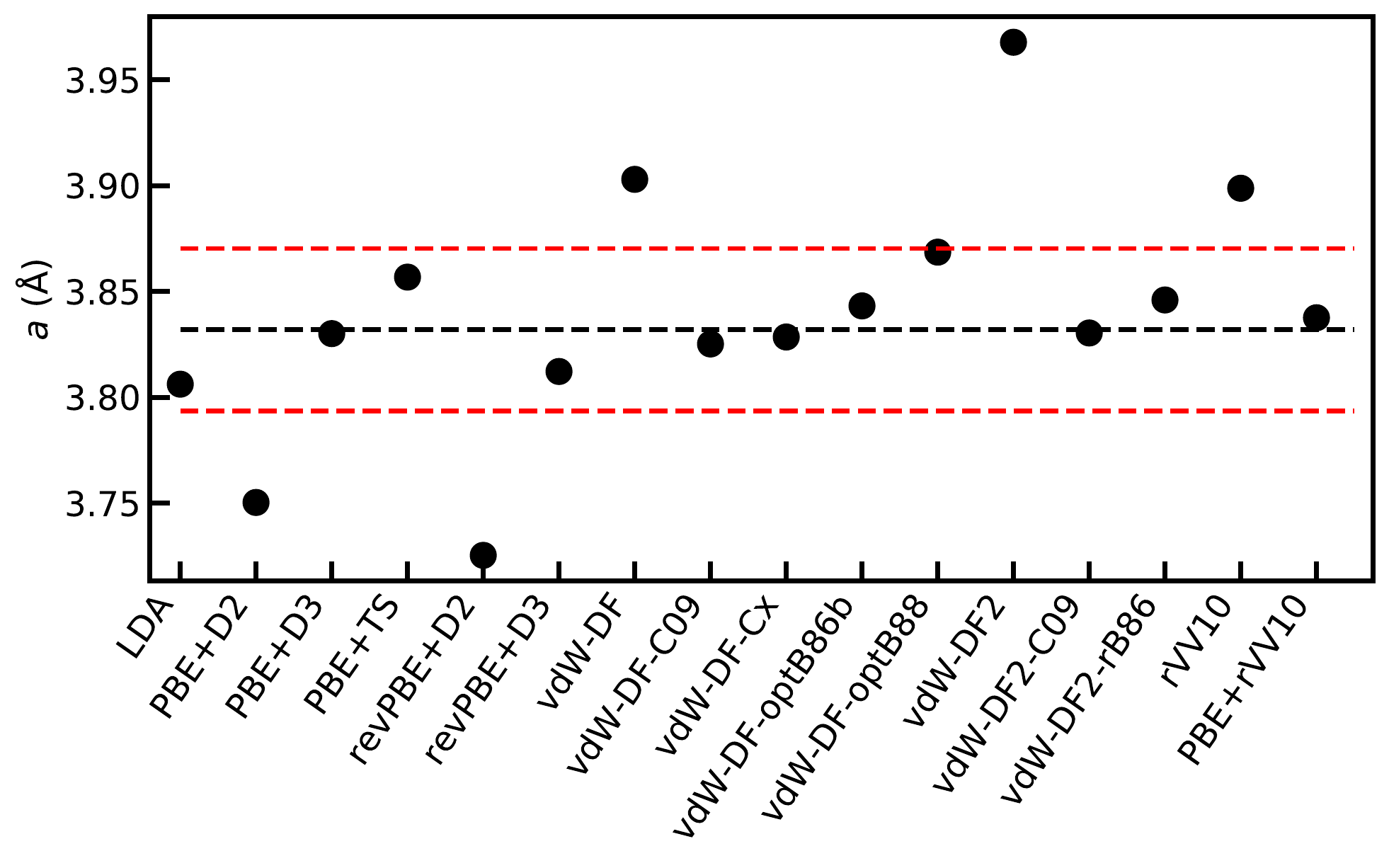}
\caption{\textbf{Lattice constant $a$ from DFT calculations with different treatments of the XC.} The calculated values are compared to the experimental value $a_\mathrm{exp}$ -- black dashed line -- and variations of $\pm{}1$~\% into it -- red dashed lines. $a_\mathrm{exp}$ has been modified to exclude the zero-point vibrational effects\cite{Hao_2012_a,Hao_2012_b}.}
\label{fig:alat}
\end{figure}

The calculated and experimental lattice constants and bulk moduli are given in
Table~\ref{tab:bulk_properties}, and the former are visualised in
Fig.~\ref{fig:alat}. The experimental value $a_\mathrm{exp}$ has been modified to exclude the
zero-point vibrational effects.\cite{Hao_2012_a,Hao_2012_b} In all calculations in the present work at $a_\mathrm{exp}$, we used the value 3.840~\AA{} that does include the
zero-point and finite temperature effects.

\begin{table}[h!]
\centering
\caption{Bulk properties of fcc-Ir with different treatments of the exchange
  and correlation in the Kohn-Sham scheme.\label{tab:bulk_properties}}\vspace{12pt}
\begin{tabular}{l|cc|cc}\hline\hline
 & \multicolumn{2}{c|}{$a$ [\AA]} & \multicolumn{2}{c}{$B$ [GPa]} \\
XC & QE & \texttt{CP2k} & QE &  \texttt{CP2k}\\\hline
LDA            & 3.806 & 3.774 & 275 & 339 \\
\hline%
PBE            & 3.857 & 3.864 & 237 & 228 \\
PBE+D2         & 3.750 & 3.752 & 311 & 314 \\
PBE+D3         & 3.830 & 3.839 & 254 & 242 \\
PBE+TS         & 3.858 &       & 239 &     \\
revPBE         & 3.868 & 3.874 & 229 & 221 \\
revPBE+D2      & 3.725 & 3.728 & 333 & 344 \\
revPBE+D3      & 3.812 & 3.819 & 257 & 242 \\
\hline%
vdW-DF         & 3.903 & 3.908 & 209 & 200 \\
vdW-DF-optB88  & 3.869 & 3.873 & 233 & 235 \\
vdW-DF-optB86b & 3.843 &       & 249 &     \\
vdW-DF-rB86    & 3.841 & 3.851 & 250 & 242 \\
vdW-DF-C09     & 3.825 &       & 265 &     \\
vdW-DF-Cx      & 3.829 &       & 262 &     \\
vdW-DF2        & 3.968 & 3.978 & 179 & 150 \\
vdW-DF2-C09    & 3.830 & 3.837 & 259 & 270 \\
vdW-DF2-rB86   & 3.846 & 3.847 & 247 & 250 \\
BEEF-DF2       &       & 3.880 &     & 228 \\
rVV10          & 3.899 & 3.905 & 217 & 209 \\
PBE+rVV10      & 3.838 & 3.848 & 253 & 245 \\
\hline%
Experiment           & \multicolumn{2}{c|}{3.832} &\multicolumn{2}{c}{320}\\\hline\hline
\end{tabular}
\end{table}


%
%
\section{Details of the calculated structures}\label{sec:structure_details}

\begin{figure*}[t!h]
\centering
\includegraphics[width=.95\textwidth]{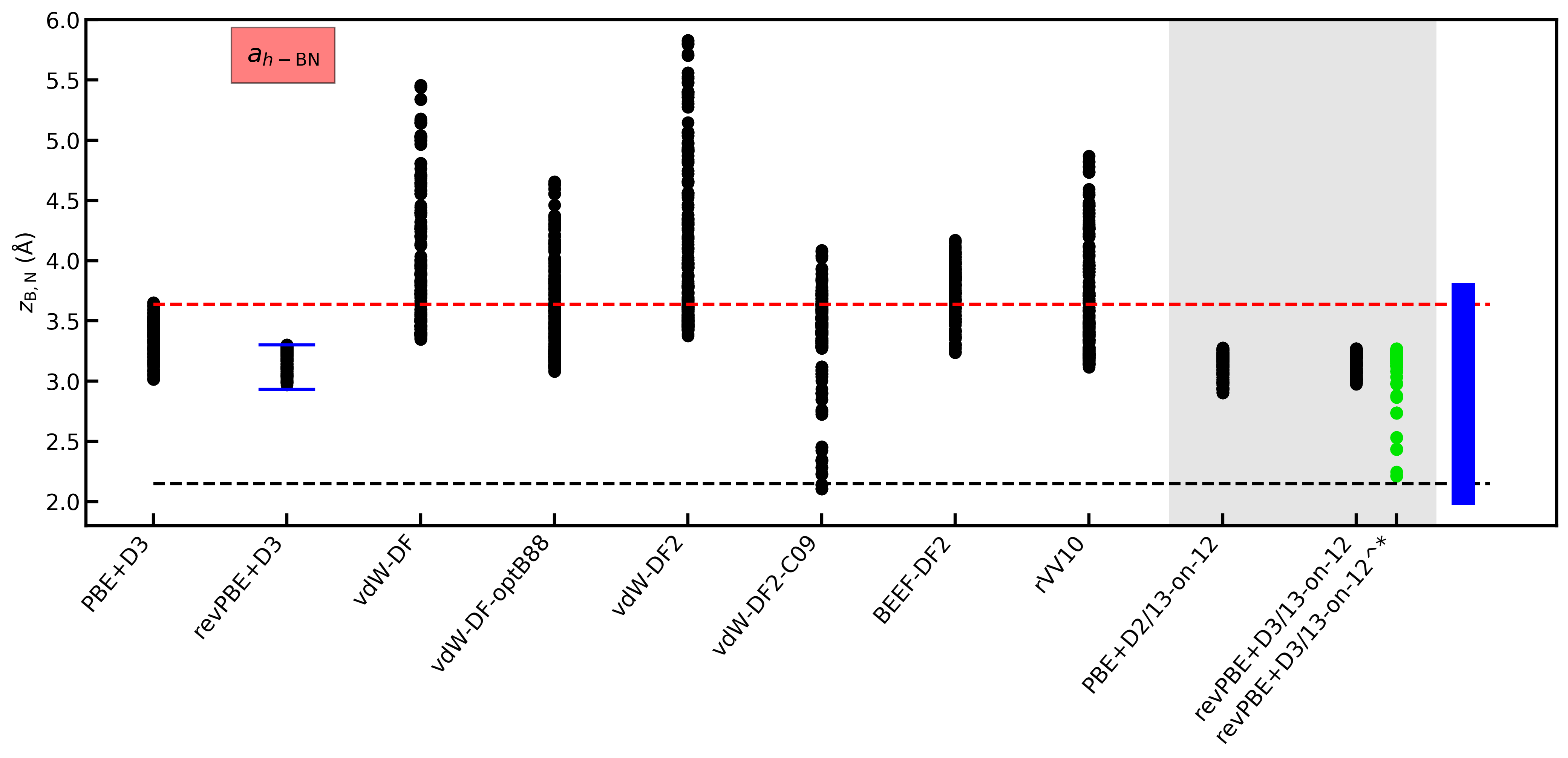}
\caption{\textbf{Height of B and N atoms in 12-on-11 structure above the average top-most layer of the substrate.} DFT-derived lattice constant of \textit{h}-BN from Ref.~\citenum{GomezDiaz_2013_a} is used. The result by Farwick zum Hagen \textit{et alia}\/ is marked with the dashed lines, minimum and maximum; green lines, values from Ref.~\citenum{Schulz_2014_a} in 13-on-12 structure using revPBE+D3 approximation.}
\label{fig:hBN_Moire_height_alat_hBN}
\end{figure*}

In Tables~\ref{tab:hBN_1x1_properties}, \ref{tab:gr_1x1_properties}, \ref{tab:hbn_nanomesh_12on11}, \ref{table:hBN_13on12_structure} and \ref{table:gr_Ir111_geometries} and Figure~\ref{fig:hBN_Moire_height_alat_hBN} we present detailed information about the structures. They contain more results (in Figure~\ref{fig:hBN_Moire_height_alat_hBN} the corrugation obtained with the lattice constant of 3.801~\AA) and further geometrical parameters. In Table~\ref{table:Moire_DFT_results} we collect the previous and present DFT results on the moir\'{e} geometries.

\begin{sidewaystable}
\centering
\caption{Adsorption properties of commensurate (1$\times$1) \textit{h}-BN on Ir(111) with different treatments of the XC in the KS scheme; $E_b$ is the binding energy, $\Delta_\mathrm{B-N}$ the geometric corrugation between the B and N and $z_\mathrm{B,N}$ the heights of the atoms above the substrate layer.\label{tab:hBN_1x1_properties}}\vspace{12pt}
\begin{tiny}
\begin{tabular}{l|ll|cccccc|cccccc}\hline\hline
 & & & \multicolumn{6}{c|}{Lateral adsorption arrangement; $E_b$/$z_\mathrm{B}$} & \multicolumn{6}{c}{Lateral adsorption arrangement; $\Delta_\mathrm{B-N}$/$z_\mathrm{N}$}\\
XC & \multicolumn{2}{c|}{Quantity} & B$_\mathrm{o}$-N$_\mathrm{h}$ & B$_\mathrm{o}$-N$_\mathrm{f}$ & B$_\mathrm{h}$-N$_\mathrm{f}$ & B$_\mathrm{h}$-N$_\mathrm{o}$ & B$_\mathrm{f}$-N$_\mathrm{o}$ & B$_\mathrm{f}$-N$_\mathrm{h}$ & B$_\mathrm{o}$-N$_\mathrm{h}$ & B$_\mathrm{o}$-N$_\mathrm{f}$ & B$_\mathrm{h}$-N$_\mathrm{f}$ & B$_\mathrm{h}$-N$_\mathrm{o}$ & B$_\mathrm{f}$-N$_\mathrm{o}$ & B$_\mathrm{f}$-N$_\mathrm{h}$\\\hline
LDA             & $E_b$ & $\Delta_\mathrm{B-N}$ & -0.533 & -0.489 & -0.090 & -0.073 & -0.088 & -0.074 & -0.140 & -0.142 & -0.020 & -0.009 & -0.018 & -0.009\\
 & $z_\mathrm{B}$ & $z_\mathrm{N}$ &  2.054 &  2.057 &  3.213 &  3.464 &  3.251 &  3.461 &  2.194 &  2.200 &  3.233 &  3.473 &  3.269 &  3.470\\
PBE+D2          & $E_b$ & $\Delta_\mathrm{B-N}$ & -0.774 & -0.730 & -0.452 & -0.446 & -0.443 & -0.447 & -0.157 & -0.158 & -0.073 & -0.022 & -0.059 & -0.022\\
 & $z_\mathrm{B}$ & $z_\mathrm{N}$ &  2.064 &  2.069 &  2.738 &  3.078 &  2.816 &  3.077 &  2.221 &  2.227 &  2.811 &  3.101 &  2.875 &  3.100\\
PBE+D3          & $E_b$ & $\Delta_\mathrm{B-N}$ & -0.368 & -0.327 & -0.151 & -0.145 & -0.149 & -0.145 & -0.149 & -0.149 & -0.013 & -0.011 & -0.013 & -0.011\\
 & $z_\mathrm{B}$ & $z_\mathrm{N}$ &  2.106 &  2.115 &  3.398 &  3.438 &  3.411 &  3.436 &  2.256 &  2.264 &  3.411 &  3.449 &  3.424 &  3.446\\
PBE+TS          & $E_b$ & $\Delta_\mathrm{B-N}$ & -0.229 & -0.193 & -0.105 & -0.098 & -0.104 & -0.098 & -0.144 & -0.143 & -0.012 & -0.009 & -0.012 & -0.010\\
 & $z_\mathrm{B}$ & $z_\mathrm{N}$ &  2.161 &  2.173 &  3.435 &  3.443 &  3.438 &  3.442 &  2.305 &  2.316 &  3.447 &  3.452 &  3.449 &  3.452\\
PBE+TS 120 Ry   & $E_b$ & $\Delta_\mathrm{B-N}$ & -0.231 & -0.197 & -0.106 & -0.100 & -0.105 & -0.100 & -0.142 & -0.141 & -0.012 & -0.009 & -0.012 & -0.009\\
 & $z_\mathrm{B}$ & $z_\mathrm{N}$ &  2.162 &  2.170 &  3.419 &  3.485 &  3.438 &  3.485 &  2.304 &  2.312 &  3.431 &  3.495 &  3.450 &  3.495\\
revPBE+D2       & $E_b$ & $\Delta_\mathrm{B-N}$ & -0.865 & -0.821 & -0.562 & -0.560 & -0.550 & -0.561 & -0.165 & -0.167 & -0.095 & -0.025 & -0.081 & -0.025\\
 & $z_\mathrm{B}$ & $z_\mathrm{N}$ &  2.056 &  2.059 &  2.650 &  3.050 &  2.720 &  3.049 &  2.221 &  2.227 &  2.745 &  3.075 &  2.801 &  3.073\\
revPBE+D3       & $E_b$ & $\Delta_\mathrm{B-N}$ & -0.471 & -0.430 & -0.233 & -0.231 & -0.229 & -0.231 & -0.156 & -0.156 & -0.031 & -0.018 & -0.028 & -0.018\\
 & $z_\mathrm{B}$ & $z_\mathrm{N}$ &  2.097 &  2.105 &  3.067 &  3.214 &  3.101 &  3.213 &  2.253 &  2.261 &  3.098 &  3.232 &  3.129 &  3.230\\
vdW-DF          & $E_b$ & $\Delta_\mathrm{B-N}$ & -0.129 & -0.128 & -0.113 & -0.111 & -0.112 & -0.110 & -0.004 & -0.004 & -0.002 & -0.001 & -0.002 & -0.001\\
 & $z_\mathrm{B}$ & $z_\mathrm{N}$ &  3.635 &  3.651 &  3.886 &  3.930 &  3.896 &  3.932 &  3.639 &  3.654 &  3.888 &  3.931 &  3.898 &  3.933\\
vdW-DF-C09      & $E_b$ & $\Delta_\mathrm{B-N}$ & -0.554 & -0.513 & -0.200 & -0.183 & -0.198 & -0.183 & -0.146 & -0.147 & -0.022 & -0.011 & -0.020 & -0.011\\
 & $z_\mathrm{B}$ & $z_\mathrm{N}$ &  2.091 &  2.097 &  3.199 &  3.390 &  3.228 &  3.387 &  2.237 &  2.244 &  3.221 &  3.401 &  3.248 &  3.398\\
vdW-DF-Cx       & $E_b$ & $\Delta_\mathrm{B-N}$ & -0.502 & -0.462 & -0.173 & -0.159 & -0.171 & -0.159 & -0.147 & -0.148 & -0.019 & -0.011 & -0.018 & -0.010\\
 & $z_\mathrm{B}$ & $z_\mathrm{N}$ &  2.097 &  2.104 &  3.268 &  3.434 &  3.291 &  3.433 &  2.244 &  2.252 &  3.287 &  3.445 &  3.309 &  3.444\\
vdW-DF-optB86b  & $E_b$ & $\Delta_\mathrm{B-N}$ & -0.404 & -0.364 & -0.172 & -0.161 & -0.170 & -0.161 & -0.138 & -0.137 & -0.013 & -0.007 & -0.012 & -0.007\\
 & $z_\mathrm{B}$ & $z_\mathrm{N}$ &  2.129 &  2.140 &  3.395 &  3.514 &  3.410 &  3.512 &  2.267 &  2.277 &  3.407 &  3.522 &  3.423 &  3.520\\
vdW-DF-optB88   & $E_b$ & $\Delta_\mathrm{B-N}$ & -0.248 & -0.213 & -0.163 & -0.155 & -0.162 & -0.155 & -0.115 & -0.111 & -0.009 & -0.005 & -0.009 & -0.005\\
 & $z_\mathrm{B}$ & $z_\mathrm{N}$ &  2.213 &  2.239 &  3.499 &  3.589 &  3.513 &  3.593 &  2.328 &  2.350 &  3.508 &  3.594 &  3.522 &  3.599\\
vdW-DF-optB88-7lrs & $E_b$ & $\Delta_\mathrm{B-N}$ & -0.278 & -0.238 & -0.165 & -0.157 & -0.164 & -0.157 & -0.116 & -0.115 & -0.009 & -0.005 & -0.009 & -0.005\\
 & $z_\mathrm{B}$ & $z_\mathrm{N}$ &  2.202 &  2.222 &  3.500 &  3.591 &  3.508 &  3.592 &  2.318 &  2.337 &  3.509 &  3.596 &  3.517 &  3.598\\
vdW-DF-optB88-comm & $E_b$ & $\Delta_\mathrm{B-N}$ & -0.156 & -0.155 & -0.142 & -0.138 & -0.140 & -0.137 & -0.005 & -0.005 & -0.005 & -0.003 & -0.004 & -0.003\\
 & $z_\mathrm{B}$ & $z_\mathrm{N}$ &  3.514 &  3.527 &  3.631 &  3.663 &  3.646 &  3.674 &  3.519 &  3.531 &  3.636 &  3.666 &  3.650 &  3.677\\
vdW-DF2         & $E_b$ & $\Delta_\mathrm{B-N}$ & -0.126 & -0.126 & -0.111 & -0.107 & -0.110 & -0.107 & -0.003 & -0.002 & -0.003 & -0.002 & -0.002 & -0.001\\
 & $z_\mathrm{B}$ & $z_\mathrm{N}$ &  3.690 &  3.727 &  3.849 &  3.913 &  3.897 &  3.915 &  3.692 &  3.729 &  3.852 &  3.915 &  3.899 &  3.916\\
vdW-DF2-C09     & $E_b$ & $\Delta_\mathrm{B-N}$ & -0.463 & -0.422 & -0.150 & -0.137 & -0.148 & -0.137 & -0.144 & -0.144 & -0.017 & -0.010 & -0.016 & -0.010\\
 & $z_\mathrm{B}$ & $z_\mathrm{N}$ &  2.101 &  2.108 &  3.304 &  3.461 &  3.326 &  3.459 &  2.245 &  2.252 &  3.321 &  3.471 &  3.342 &  3.468\\
vdW-DF2-rB86    & $E_b$ & $\Delta_\mathrm{B-N}$ & -0.355 & -0.315 & -0.147 & -0.137 & -0.145 & -0.137 & -0.133 & -0.133 & -0.012 & -0.007 & -0.011 & -0.007\\
 & $z_\mathrm{B}$ & $z_\mathrm{N}$ &  2.138 &  2.150 &  3.433 &  3.547 &  3.448 &  3.546 &  2.272 &  2.283 &  3.444 &  3.554 &  3.459 &  3.553\\
vdW-DF2-rB86 08kpt & $E_b$ & $\Delta_\mathrm{B-N}$ & -0.328 & -0.196 & -0.150 & -0.139 & -0.147 & -0.138 & -0.135 & -0.031 & -0.013 & -0.007 & -0.012 & -0.007\\
 & $z_\mathrm{B}$ & $z_\mathrm{N}$ &  2.125 &  2.869 &  3.391 &  3.528 &  3.423 &  3.533 &  2.260 &  2.900 &  3.405 &  3.536 &  3.435 &  3.540\\
rVV10           & $E_b$ & $\Delta_\mathrm{B-N}$ & -0.225 & -0.222 & -0.184 & -0.176 & -0.183 & -0.176 & -0.012 & -0.011 & -0.009 & -0.006 & -0.009 & -0.006\\
 & $z_\mathrm{B}$ & $z_\mathrm{N}$ &  3.262 &  3.304 &  3.552 &  3.626 &  3.559 &  3.627 &  3.274 &  3.314 &  3.561 &  3.632 &  3.567 &  3.633\\
PBE+rVV10       & $E_b$ & $\Delta_\mathrm{B-N}$ & -0.495 & -0.454 & -0.203 & -0.190 & -0.201 & -0.190 & -0.149 & -0.149 & -0.017 & -0.010 & -0.016 & -0.010\\
 & $z_\mathrm{B}$ & $z_\mathrm{N}$ &  2.105 &  2.113 &  3.320 &  3.457 &  3.337 &  3.454 &  2.254 &  2.262 &  3.337 &  3.468 &  3.354 &  3.464\\
\hline\hline
\end{tabular}
\end{tiny}
\end{sidewaystable}

\begin{table*}[ht]
\centering
\caption{Adsorption properties of commensurate (1$\times$1) gr on Ir(111) with different treatments of the XC in the KS scheme; $E_b$ is the binding energy, $\Delta_\mathrm{C-C}$ the geometric corrugation between the two carbon atoms C$_1$ and C$_2$ and $z_\mathrm{C_1,C_2}$ the heights of the atoms above the substrate layer.\label{tab:gr_1x1_properties}}\vspace{12pt}
  \renewcommand{\arraystretch}{0.75}
\begin{tabular}{l|ll|ccc|ccc}\hline\hline
 & & & \multicolumn{6}{c}{Lateral adsorption arrangement}\\
 & & & \multicolumn{3}{c|}{$E_b$/$z_\mathrm{C-C}$} & \multicolumn{3}{c}{$\Delta_\mathrm{CC}$/$z_\mathrm{C_1}$}\\
XC & \multicolumn{2}{c|}{Quantity} & C$_\mathrm{fo}$ & C$_\mathrm{ho}$ & C$_\mathrm{hf}$  & C$_\mathrm{fo}$ & C$_\mathrm{ho}$ & C$_\mathrm{hf}$\\\hline
LDA             & $E_b$ & $\Delta_\mathrm{C-C}$ & -0.509 & -0.433 & -0.059 & -0.040 &  0.053 &  0.000\\
 & $z_\mathrm{C_1}$ & $z_\mathrm{C_2}$ &  2.218 &  2.209 &  3.639 &  2.177 &  2.262 &  3.639\\
PBE+D2          & $E_b$ & $\Delta_\mathrm{C-C}$ & -0.707 & -0.639 & -0.355 & -0.054 &  0.066 &  0.000\\
 & $z_\mathrm{C_1}$ & $z_\mathrm{C_2}$ &  2.272 &  2.253 &  3.168 &  2.218 &  2.319 &  3.169\\
PBE+D3          & $E_b$ & $\Delta_\mathrm{C-C}$ & -0.271 & -0.148 & -0.116 & -0.057 &  0.005 &  0.000\\
 & $z_\mathrm{C_1}$ & $z_\mathrm{C_2}$ &  2.331 &  3.349 &  3.675 &  2.274 &  3.354 &  3.675\\
PBE+TS          & $E_b$ & $\Delta_\mathrm{C-C}$ & -0.218 & -0.118 & -0.084 & -0.057 &  0.005 &  0.000\\
 & $z_\mathrm{C_1}$ & $z_\mathrm{C_2}$ &  2.347 &  3.332 &  3.626 &  2.291 &  3.337 &  3.626\\
revPBE+D2       & $E_b$ & $\Delta_\mathrm{C-C}$ & -0.780 & -0.711 & -0.448 & -0.058 &  0.070 &  0.000\\
 & $z_\mathrm{C_1}$ & $z_\mathrm{C_2}$ &  2.279 &  2.258 &  3.126 &  2.221 &  2.328 &  3.126\\
revPBE+D3       & $E_b$ & $\Delta_\mathrm{C-C}$ & -0.323 & -0.271 & -0.180 & -0.063 &  0.065 &  0.000\\
 & $z_\mathrm{C_1}$ & $z_\mathrm{C_2}$ &  2.374 &  2.356 &  3.421 &  2.311 &  2.421 &  3.421\\
vdW-DF          & $E_b$ & $\Delta_\mathrm{C-C}$ & -0.121 & -0.120 & -0.110 & -0.002 &  0.001 &  0.000\\
 & $z_\mathrm{C_1}$ & $z_\mathrm{C_2}$ &  3.813 &  3.851 &  4.033 &  3.812 &  3.852 &  4.033\\
vdW-DF-C09      & $E_b$ & $\Delta_\mathrm{C-C}$ & -0.539 & -0.472 & -0.168 & -0.048 &  0.059 &  0.000\\
 & $z_\mathrm{C_1}$ & $z_\mathrm{C_2}$ &  2.265 &  2.249 &  3.531 &  2.217 &  2.308 &  3.531\\
vdW-DF-Cx       & $E_b$ & $\Delta_\mathrm{C-C}$ & -0.487 & -0.422 & -0.146 & -0.049 &  0.059 &  0.000\\
 & $z_\mathrm{C_1}$ & $z_\mathrm{C_2}$ &  2.272 &  2.256 &  3.584 &  2.223 &  2.316 &  3.584\\
vdW-DF-optB86b  & $E_b$ & $\Delta_\mathrm{C-C}$ & -0.401 & -0.342 & -0.152 & -0.054 &  0.062 &  0.000\\
 & $z_\mathrm{C_1}$ & $z_\mathrm{C_2}$ &  2.302 &  2.283 &  3.647 &  2.248 &  2.346 &  3.647\\
vdW-DF-optB88   & $E_b$ & $\Delta_\mathrm{C-C}$ & -0.272 & -0.179 & -0.148 & -0.062 &  0.005 &  0.000\\
 & $z_\mathrm{C_1}$ & $z_\mathrm{C_2}$ &  2.362 &  3.385 &  3.711 &  2.299 &  3.390 &  3.711\\
vdW-DF2         & $E_b$ & $\Delta_\mathrm{C-C}$ & -0.120 & -0.118 & -0.107 & -0.002 &  0.002 &  0.000\\
 & $z_\mathrm{C_1}$ & $z_\mathrm{C_2}$ &  3.798 &  3.864 &  4.023 &  3.796 &  3.865 &  4.023\\
vdW-DF-C09      & $E_b$ & $\Delta_\mathrm{C-C}$ & -0.539 & -0.472 & -0.168 & -0.048 &  0.059 &  0.000\\
 & $z_\mathrm{C_1}$ & $z_\mathrm{C_2}$ &  2.265 &  2.249 &  3.531 &  2.217 &  2.308 &  3.531\\
vdW-DF2-rB86    & $E_b$ & $\Delta_\mathrm{C-C}$ & -0.358 & -0.300 & -0.129 & -0.055 &  0.063 &  0.000\\
 & $z_\mathrm{C_1}$ & $z_\mathrm{C_2}$ &  2.307 &  2.288 &  3.676 &  2.252 &  2.351 &  3.676\\
rVV10           & $E_b$ & $\Delta_\mathrm{C-C}$ & -0.164 & -0.199 & -0.169 & -0.059 &  0.005 &  0.000\\
 & $z_\mathrm{C_1}$ & $z_\mathrm{C_2}$ &  2.478 &  3.447 &  3.700 &  2.419 &  3.452 &  3.700\\
PBE+rVV10       & $E_b$ & $\Delta_\mathrm{C-C}$ & -0.473 & -0.411 & -0.178 & -0.049 &  0.058 &  0.000\\
 & $z_\mathrm{C_1}$ & $z_\mathrm{C_2}$ &  2.286 &  2.271 &  3.598 &  2.237 &  2.329 &  3.598\\
\hline\hline
\end{tabular}
\end{table*}

\begin{sidewaystable}
\centering
\caption{Properties of \textit{h}-BN/Ir(111) moir\'{e} structure in 12-on-11 cell from DFT calculations; alat is the lattice constant taken from free-standing \textit{h}-BN layer ("\textit{h}-BN"), optimised with the corresponding DFT-XC ("xc") or experiments ("exp"), $\Delta_\mathrm{B}$ and $\Delta_\mathrm{N}$ are the corrugation in the respective species, $\Delta_\mathrm{BN}$ the combined corrugation, $\Delta_\mathrm{Ir_1}$ the corrugation in the first substrate layer; $\Delta_\mathrm{B-N}^\mathrm{min}$ and $\Delta_\mathrm{B-N}^\mathrm{max}$ are the local corrugations, $\overline{z}_{h\mathrm{-BN}}$, $\overline{z}_\mathrm{Ir_1}$ the average heights, ${z}_{h\mathrm{-BN}}^\mathrm{min}$ the lowest atom in the \textit{h}-BN layer, ${z}_\mathrm{Ir_1}^\mathrm{max}$ the vertical coordinate of the highest substrate layer. \label{tab:hbn_nanomesh_12on11}}\vspace{12pt}
\begin{tiny}
\begin{tabular}{l|c|cccc|cc|cccc}
 & \multicolumn{1}{c|}{alat} & \multicolumn{4}{c|}{corrugation} & \multicolumn{2}{c|}{local} & \multicolumn{4}{c}{distance} \\
& \multicolumn{1}{c|}{$a$} & $\Delta_\mathrm{B}$ & $\Delta_\mathrm{N}$ & $\Delta_\mathrm{BN}$ & $\Delta_\mathrm{Ir_1}$ & $\Delta_\mathrm{B-N}^\mathrm{min}$ & $\Delta_\mathrm{B-N}^\mathrm{max}$ & $\overline{z}_{h\mathrm{-BN}}$-$\overline{z}_\mathrm{Ir_1}$ & ${z}_{h\mathrm{-BN}}^\mathrm{min}$-${z}_\mathrm{Ir_1}^\mathrm{max}$ & ${z}_{h\mathrm{-BN}}^\mathrm{min}$-$\overline{z}_\mathrm{Ir_1}$ & ${z}_{h\mathrm{-BN}}^\mathrm{max}$-$\overline{z}_\mathrm{Ir_1}$ \\\hline\hline
PBE+D3        & \textit{h}-BN & 0.628 & 0.632 & 0.637 & 0.034 & -0.004 & -0.009 & 3.372 & 3.003 & 3.015 & 3.652\\
PBE+D3$^*$    & \textit{h}-BN & 1.853 & 1.953 & 1.953 & 0.110 & 0.064 & -0.037 & 3.364 & 1.980 & 1.999 & 3.952\\
revPBE+D3     & \textit{h}-BN & 0.317 & 0.316 & 0.333 & 0.031 & -0.017 & -0.015 & 3.192 & 2.958 & 2.970 & 3.302\\
vdW-DF        & \textit{h}-BN & 2.081 & 2.109 & 2.109 & 0.030 & 0.012 & -0.016 & 4.032 & 3.336 & 3.348 & 5.457\\
vdW-DF-optB88 & \textit{h}-BN & 1.531 & 1.574 & 1.574 & 0.040 & 0.024 & -0.018 & 3.582 & 3.066 & 3.083 & 4.657\\
vdW-DF2       & \textit{h}-BN & 2.397 & 2.453 & 2.453 & 0.040 & 0.044 & -0.012 & 4.148 & 3.361 & 3.376 & 5.829\\
vdW-DF2-C09   & \textit{h}-BN & 1.937 & 1.861 & 1.982 & 0.047 & -0.121 & -0.045 & 3.365 & 2.080 & 2.105 & 4.086\\
BEEF-DF2      & \textit{h}-BN & 0.919 & 0.936 & 0.936 & 0.041 & 0.006 & -0.010 & 3.723 & 3.230 & 3.237 & 4.172\\
rVV10         & \textit{h}-BN & 1.684 & 1.752 & 1.752 & 0.047 & 0.020 & -0.048 & 3.649 & 3.095 & 3.116 & 4.868\\
\hline
LDA           & xc   & 3.323 & 3.356 & 3.498 & 0.127 & -0.142 & -0.175 & 3.565 & 1.952 & 2.046 & 5.543\\
PBE+D2        & xc   & 0.212 & 0.214 & 0.229 & 0.028 & -0.015 & -0.017 & 3.090 & 2.914 & 2.923 & 3.152\\
PBE+D2$^*$    & xc   & 1.073 & 0.993 & 1.093 & 0.055 & -0.100 & -0.020 & 3.061 & 2.037 & 2.059 & 3.152\\
PBE+D3        & xc   & 0.466 & 0.466 & 0.474 & 0.023 & -0.009 & -0.009 & 3.353 & 3.014 & 3.021 & 3.495\\
PBE+D3$^*$    & xc   & 1.279 & 1.256 & 1.289 & 0.054 & -0.033 & -0.010 & 3.326 & 2.165 & 2.191 & 3.480\\
revPBE+D3     & xc   & 0.313 & 0.312 & 0.326 & 0.028 & -0.015 & -0.013 & 3.173 & 2.959 & 2.967 & 3.294\\
revPBE+D3$^*$ & xc   & 0.308 & 0.314 & 0.324 & 0.028 & -0.010 & -0.016 & 3.175 & 2.960 & 2.968 & 3.292\\
vdW-DF        & xc   & 0.312 & 0.315 & 0.315 & 0.005 & 0.002 & -0.001 & 3.759 & 3.556 & 3.557 & 3.872\\
vdW-DF-optB88 & xc   & 0.553 & 0.560 & 0.560 & 0.017 & 0.001 & -0.006 & 3.450 & 3.075 & 3.080 & 3.640\\
vdW-DF-optB88$^*$ &xc& 0.556 & 0.567 & 0.567 & 0.017 & 0.004 & -0.008 & 3.450 & 3.071 & 3.076 & 3.643\\
vdW-DF-rB86   & xc   & 1.356 & 1.284 & 1.363 & 0.034 & -0.079 & -0.008 & 3.318 & 2.187 & 2.206 & 3.569\\
vdW-DF2       & xc   & 0.215 & 0.219 & 0.219 & 0.003 & 0.001 & -0.002 & 3.751 & 3.616 & 3.617 & 3.836\\
vdW-DF2-C09   & xc   & 1.495 & 1.380 & 1.505 & 0.048 & -0.125 & -0.010 & 3.304 & 2.072 & 2.100 & 3.604\\
vdW-DF2-rB86  & xc   & 1.310 & 1.291 & 1.318 & 0.041 & -0.027 & -0.008 & 3.389 & 2.294 & 2.317 & 3.634\\
vdW-DF2-rB86 7lrs & xc   & 1.492 & 1.389 & 1.498 & 0.066 & -0.109 & -0.006 & 3.377 & 2.087 & 2.126 & 3.624\\
vdW-DF2-rB86 s1 & xc   & 1.445 & 1.414 & 1.455 & 0.045 & -0.041 & -0.010 & 3.379 & 2.161 & 2.185 & 3.640\\
vdW-DF2-rB86 s1 7lrs & xc   & 1.512 & 1.451 & 1.524 & 0.052 & -0.073 & -0.012 & 3.383 & 2.110 & 2.141 & 3.665\\
vdW-DF2-rB86 s2 & xc   & 1.404 & 1.387 & 1.414 & 0.042 & -0.027 & -0.010 & 3.383 & 2.200 & 2.222 & 3.636\\
BEEF-DF2      & xc   & 0.481 & 0.488 & 0.488 & 0.005 & 0.002 & -0.004 & 3.714 & 3.398 & 3.400 & 3.888\\
BEEF-DF2$^*$  &xc & 0.462 & 0.486 & 0.486 & 0.005 & 0.012 & -0.011 & 3.726 & 3.400 & 3.402 & 3.888\\
rVV10         & xc   & 0.412 & 0.418 & 0.419 & 0.014 & -0.001 & -0.007 & 3.482 & 3.211 & 3.215 & 3.634\\
PBE+rVV10     & xc   & 1.428 & 1.308 & 1.436 & 0.047 & -0.129 & -0.009 & 3.294 & 2.075 & 2.101 & 3.538\\
\hline
LDA           & exp & 1.886 & 1.730 & 1.900 & 0.068 & -0.170 & -0.013 & 3.179 & 1.953 & 2.004 & 3.903\\
PBE+D2        & exp & 0.235 & 0.237 & 0.253 & 0.028 & -0.016 & -0.018 & 3.082 & 2.889 & 2.898 & 3.150\\
PBE+D2$^*$    & exp & 1.087 & 0.955 & 1.104 & 0.057 & -0.149 & -0.017 & 3.024 & 2.015 & 2.041 & 3.145\\
PBE+D3        & exp & 0.464 & 0.464 & 0.473 & 0.023 & -0.009 & -0.009 & 3.353 & 3.015 & 3.021 & 3.494\\
PBE+D3$^*$    & exp & 1.280 & 1.256 & 1.290 & 0.054 & -0.034 & -0.010 & 3.325 & 2.164 & 2.191 & 3.481\\
revPBE+D3     & exp & 0.281 & 0.278 & 0.293 & 0.027 & -0.015 & -0.013 & 3.169 & 2.959 & 2.968 & 3.261\\
revPBE+D3$^*$ & exp & 0.280 & 0.281 & 0.293 & 0.027 & -0.012 & -0.014 & 3.169 & 2.959 & 2.968 & 3.262\\
vdW-DF        & exp & 0.520 & 0.520 & 0.522 & 0.015 & -0.002 & -0.002 & 3.779 & 3.390 & 3.393 & 3.915\\
vdW-DF-optB88 & exp & 0.653 & 0.662 & 0.662 & 0.024 & 0.003 & -0.006 & 3.468 & 3.021 & 3.027 & 3.689\\
vdW-DF-rB86   & exp & 1.367 & 1.294 & 1.375 & 0.033 & -0.081 & -0.008 & 3.330 & 2.184 & 2.200 & 3.575\\
vdW-DF2       & exp & 0.508 & 0.518 & 0.518 & 0.024 & 0.005 & -0.005 & 3.783 & 3.478 & 3.486 & 4.004\\
vdW-DF2-C09   & exp & 1.440 & 1.322 & 1.449 & 0.050 & -0.127 & -0.009 & 3.291 & 2.068 & 2.097 & 3.546\\
vdW-DF2-rB86  & exp & 1.328 & 1.305 & 1.335 & 0.038 & -0.030 & -0.007 & 3.396 & 2.285 & 2.305 & 3.640\\
BEEF-DF2      & exp & 0.554 & 0.562 & 0.562 & 0.042 & 0.006 & -0.002 & 3.673 & 3.298 & 3.320 & 3.882\\
rVV10         & exp & 0.556 & 0.564 & 0.564 & 0.026 & 0.001 & -0.008 & 3.506 & 3.138 & 3.145 & 3.710\\
PBE+rVV10     & exp & 1.420 & 1.303 & 1.429 & 0.046 & -0.126 & -0.009 & 3.303 & 2.080 & 2.105 & 3.533\\
\hline\hline
\end{tabular}
\end{tiny}
\end{sidewaystable}

\begin{sidewaystable}
\centering
\caption{Properties of \textit{h}-BN/Ir(111) moir\'{e} structure in 13-on-12 cell from DFT calculations; symbols as in previous caption.\label{table:hBN_13on12_structure}}\vspace{12pt}
\begin{tabular}{l|c|cccc|cc|cccc}
 & \multicolumn{1}{c|}{alat} & \multicolumn{4}{c|}{corrugation} & \multicolumn{2}{c|}{local} & \multicolumn{4}{c}{distance} \\
& \multicolumn{1}{c|}{$a$} & $\Delta_\mathrm{B}$ & $\Delta_\mathrm{N}$ & $\Delta_\mathrm{BN}$ & $\Delta_\mathrm{Ir_1}$ & $\Delta_\mathrm{B-N}^\mathrm{min}$ & $\Delta_\mathrm{B-N}^\mathrm{max}$ & $\overline{z}_{h\mathrm{-BN}}$-$\overline{z}_\mathrm{Ir_1}$ & ${z}_{h\mathrm{-BN}}^\mathrm{min}$-${z}_\mathrm{Ir_1}^\mathrm{max}$ & ${z}_{h\mathrm{-BN}}^\mathrm{min}$-$\overline{z}_\mathrm{Ir_1}$ & ${z}_{h\mathrm{-BN}}^\mathrm{max}$-$\overline{z}_\mathrm{Ir_1}$ \\\hline\hline
PBE+D2/13-on-12 & \textit{h}-BN & 0.361 & 0.365 & 0.372 & 0.026 & -0.007 & -0.011 & 3.151 & 2.896 & 2.905 & 3.277\\
revPBE+D3/13-on-12 & \textit{h}-BN & 0.282 & 0.278 & 0.296 & 0.025 & -0.018 & -0.014 & 3.174 & 2.967 & 2.975 & 3.271\\
revPBE+D3/13-on-12$^*$ & \textit{h}-BN & 1.045 & 1.023 & 1.058 & 0.047 & -0.034 & -0.013 & 3.153 & 2.193 & 2.213 & 3.271\\
\hline
PBE+D2/13-on-12 & xc   & 0.206 & 0.208 & 0.223 & 0.028 & -0.016 & -0.017 & 3.099 & 2.930 & 2.937 & 3.160\\
PBE+D2/13-on-12$^*$ & xc   & 1.128 & 1.027 & 1.139 & 0.049 & -0.112 & -0.011 & 3.080 & 2.035 & 2.054 & 3.192\\
vdW-DF2-rB86/13-on-12 & xc   & 0.754 & 0.779 & 0.779 & 0.015 & 0.017 & -0.007 & 3.391 & 2.837 & 2.841 & 3.621\\
vdW-DF2-rB86/13-on-12$^*$ & xc   & 1.374 & 1.339 & 1.381 & 0.047 & -0.042 & -0.007 & 3.365 & 2.200 & 2.231 & 3.612\\
\hline
vdW-DF2-rB86/13-on-12 & exp & 0.780 & 0.813 & 0.813 & 0.016 & 0.025 & -0.008 & 3.396 & 2.815 & 2.820 & 3.633\\
vdW-DF2-rB86/13-on-12$^*$ & exp & 1.382 & 1.351 & 1.391 & 0.046 & -0.040 & -0.009 & 3.375 & 2.202 & 2.231 & 3.622\\
\hline\hline
\end{tabular}
\end{sidewaystable}

\begin{table*}[ht]
\centering
\caption{Properties of gr/Ir(111) moir\'{e} structure in 10-on-9 cell from DFT calculations; alat is the lattice constant taken either the DFT with the approximation to the exchange-correlation ("xc") or experiments ("exp"), $\Delta_\mathrm{CC}$ the maximal corrugation in the graphene layer, $\Delta_\mathrm{Ir_1}$ the corrugation in the first substrate layer; $\overline{z}_{\mathrm{CC}}$, $\overline{z}_\mathrm{Ir_1}$ the average heights, ${z}_{\mathrm{gr}}^\mathrm{min}$ the lowest atom in the graphene layer, ${z}_\mathrm{Ir_1}^\mathrm{max}$ the vertical coordinate of the highest substrate layer.\label{table:gr_Ir111_geometries}}\vspace{12pt}
\begin{footnotesize}
  \renewcommand{\arraystretch}{0.73}
\begin{tabular}{l|c|cc|cccc}
 & \multicolumn{1}{c|}{alat} & \multicolumn{2}{c|}{corrugation} & \multicolumn{4}{c}{distance} \\
& \multicolumn{1}{c|}{$a$} & $\Delta_\mathrm{CC}$ & $\Delta_\mathrm{Ir_1}$ & $\overline{z}_{\mathrm{gr}}$-$\overline{z}_\mathrm{Ir_1}$ & ${z}_{\mathrm{gr}}^\mathrm{min}$-${z}_\mathrm{Ir_1}^\mathrm{max}$ & ${z}_{\mathrm{gr}}^\mathrm{min}$-$\overline{z}_\mathrm{Ir_1}$ & ${z}_{\mathrm{gr}}^\mathrm{max}$-$\overline{z}_\mathrm{Ir_1}$ \\\hline\hline
LDA           & xc   & 3.029 & 0.075 & 3.600 & 2.070 & 2.103 & 5.132\\
PBE+D3        & xc   & 0.372 & 0.014 & 3.471 & 3.314 & 3.322 & 3.695\\
revPBE+D3     & xc   & 0.296 & 0.020 & 3.317 & 3.173 & 3.185 & 3.481\\
vdW-DF        & xc   & 0.175 & 0.002 & 3.796 & 3.730 & 3.731 & 3.906\\
vdW-DF-optB88 & xc   & 0.351 & 0.009 & 3.492 & 3.348 & 3.354 & 3.704\\
vdW-DF-rB86   & xc   & 0.491 & 0.014 & 3.383 & 3.173 & 3.181 & 3.672\\
vdW-DF2       & xc   & 0.119 & 0.001 & 3.802 & 3.757 & 3.757 & 3.876\\
vdW-DF2-C09   & xc   & 0.719 & 0.019 & 3.372 & 3.057 & 3.067 & 3.787\\
vdW-DF2-rB86  & xc   & 0.515 & 0.014 & 3.439 & 3.222 & 3.230 & 3.745\\
vdW-DF2-rB86 7lrs & xc   & 0.552 & 0.024 & 3.417 & 3.168 & 3.182 & 3.735\\
BEEF-DF2      & xc   & 0.280 & 0.002 & 3.717 & 3.605 & 3.606 & 3.886\\
rVV10         & xc   & 0.262 & 0.007 & 3.530 & 3.424 & 3.428 & 3.691\\
PBE+rVV10     & xc   & 0.506 & 0.015 & 3.383 & 3.163 & 3.171 & 3.677\\
\hline
LDA           & exp & 0.826 & 0.014 & 3.357 & 2.974 & 2.982 & 3.808\\
PBE+D2 (QE)   & exp & 0.242 & 0.019 & 3.140 & 3.000 & 3.011 & 3.253\\
PBE+D3        & exp & 0.368 & 0.014 & 3.471 & 3.316 & 3.324 & 3.692\\
PBE+TS (QE)   & exp & 0.318 & 0.013 & 3.507 & 3.356 & 3.364 & 3.682\\
revPBE+D3     & exp & 0.265 & 0.016 & 3.313 & 3.190 & 3.200 & 3.464\\
vdW-DF        & exp & 0.529 & 0.012 & 3.807 & 3.604 & 3.611 & 4.140\\
vdW-DF-optB88 & exp & 0.486 & 0.015 & 3.502 & 3.300 & 3.308 & 3.794\\
vdW-DF-rB86   & exp & 0.544 & 0.017 & 3.388 & 3.153 & 3.162 & 3.705\\
vdW-DF2       & exp & 0.513 & 0.019 & 3.826 & 3.635 & 3.646 & 4.159\\
vdW-DF2-C09   & exp & 0.603 & 0.016 & 3.357 & 3.083 & 3.091 & 3.694\\
vdW-DF2-C09 (QE) & exp & 0.607 & 0.015 & 3.342 & 3.069 & 3.077 & 3.685\\
vdW-DF2-C09-ecut (QE) & exp & 0.608 & 0.015 & 3.343 & 3.070 & 3.078 & 3.686\\
vdW-DF2-C09-kpt (QE) & exp & 0.719 & 0.015 & 3.326 & 2.966 & 2.975 & 3.694\\
vdW-DF2-rB86  & exp & 0.561 & 0.016 & 3.442 & 3.206 & 3.215 & 3.775\\
vdW-DF2-rB86 (QE) & exp & 0.508 & 0.011 & 3.482 & 3.273 & 3.279 & 3.787\\
vdW-DF2-rB86-kpt (QE) & exp & 0.543 & 0.012 & 3.480 & 3.242 & 3.249 & 3.792\\
BEEF-DF2      & exp & 0.414 & 0.004 & 3.734 & 3.572 & 3.573 & 3.987\\
rVV10         & exp & 0.448 & 0.018 & 3.545 & 3.359 & 3.369 & 3.817\\
PBE+rVV10     & exp & 0.548 & 0.017 & 3.386 & 3.145 & 3.155 & 3.703\\
\hline\hline
\end{tabular}
\end{footnotesize}
\end{table*}

\begin{table*}[h!]
\centering
\caption{Structural comparison from previous and present DFT calculations on  \textit{h}-BN and gr on Ir(111).\label{table:Moire_DFT_results}}
\begin{tabular}{l|c|c|c|l}
 & \multicolumn{1}{c|}{alat} & corrugation & distance \\
& \multicolumn{1}{c|}{$a$} & $\Delta_\mathrm{BN}$ & $\overline{z}_{h\mathrm{-BN}}$-$\overline{z}_\mathrm{Ir_1}$ & Reference \\\hline\hline
12-on-11/vdW-DF2-rB86  & N/A & 1.50  & 3.24  & Ref. \citenum{zumHagen_2016_a}\\
12-on-11/vdW-DF2-rB86  & xc  & 1.318 & 3.389 & Present work\\
13-on-12/revPBE+D3     & xc  & 0.325 & 3.188 & Ref. \citenum{Schulz_2014_a}\\
13-on-12/revPBE+D3     & xc  & 0.337 & 3.187 & Ref. \citenum{Schulz_2014_a}\\
13-on-12/revPBE+D3     & xc  & 0.296 & 3.174 & Present work\\
13-on-12/revPBE+D3$^*$ & xc  & 1.058 & 3.153 & Present work\\
13-on-12/PBE+D2        & xc  & 0.223 & 3.099 & Present work\\
13-on-12/PBE+D2$^*$    & xc  & 1.058 & 3.080 & Present work\\
13-on-12/PBE+D2        & N/A & 1.4 & 3.8  & Ref. \citenum{Liu2014_NanoLett}\\
\hline\hline
& \multicolumn{1}{c|}{$a$} & $\Delta_\mathrm{CC}$ & $\overline{z}_{\mathrm{C}}$-$\overline{z}_\mathrm{Ir_1}$ & \\\hline
gr/vdW-DF(post-process)& N/A & 0.35  & 3.41  & Ref. \citenum{Busse2011_PRL}\\
gr/vdW-DF2-rB86        & N/A & 0.36  & 3.43  & Ref. \citenum{zumHagen_2016_a}\\
gr/vdW-DF2-rB86        & xc  & 0.515 & 3.439 & Present work\\
gr/PBE+D2              & N/A & 0.2   & 4.2   & Ref. \citenum{Liu2014_NanoLett}\\
gr/PBE+D2              & xc  & 0.242 & 3.140 & Present work (QE)\\%
\hline\hline
\end{tabular}
\end{table*}

\begin{figure*}[ht]
\includegraphics[width=\textwidth]{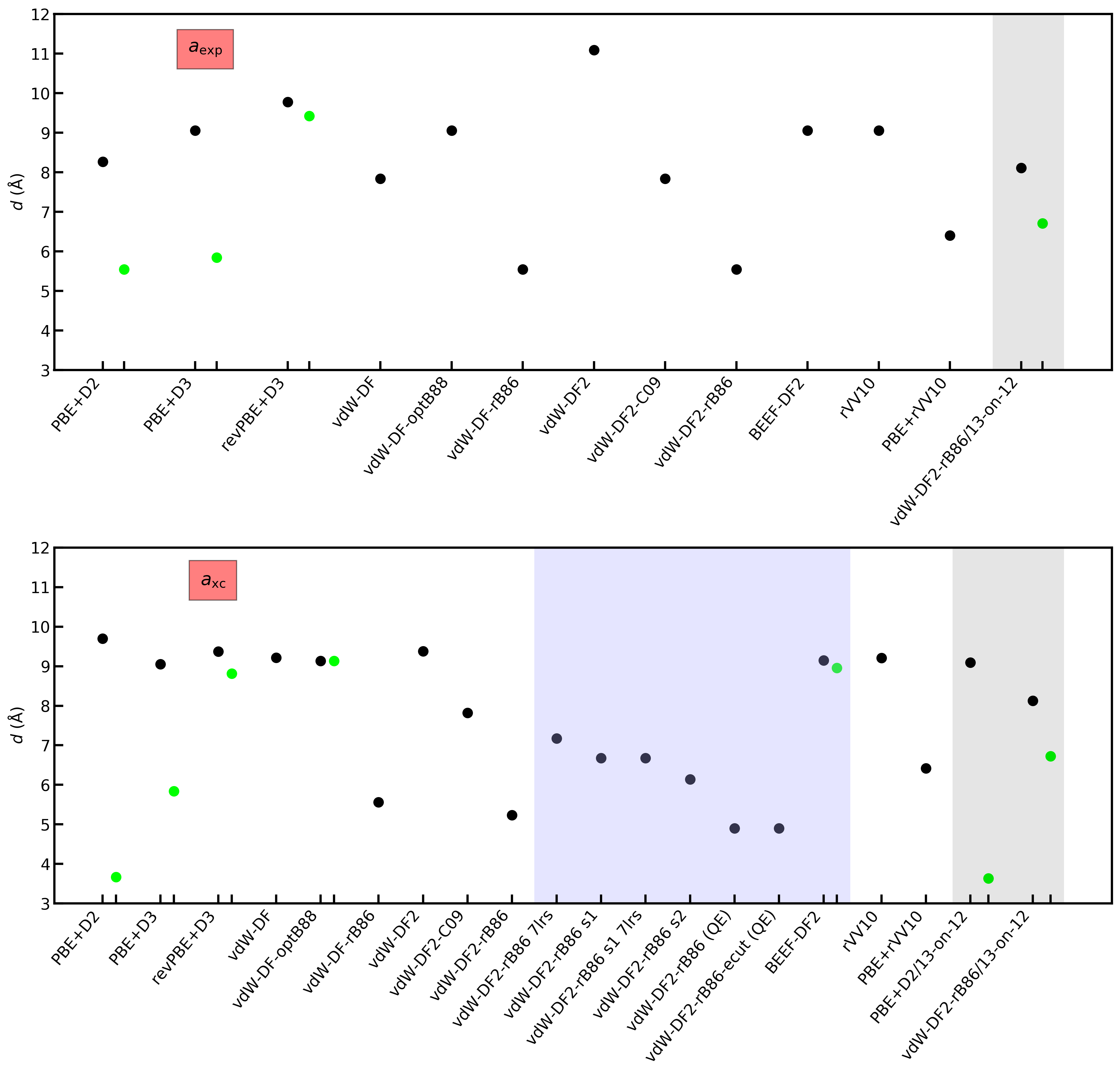}
\caption{\textbf{The apparent diameter of the depression $d$ in the \textit{h}-BN/Ir(111) structures.} Evaluated using a geometric definition for the depression and assuming a spherical shape for it; as the values using the LDA in the case of \textit{h}-BN/Ir(111) are very different and the depression is non-circular, we exclude them here.}
\label{fig:hBN_poresize}
\end{figure*}

\begin{figure*}[ht]
\includegraphics[width=\textwidth]{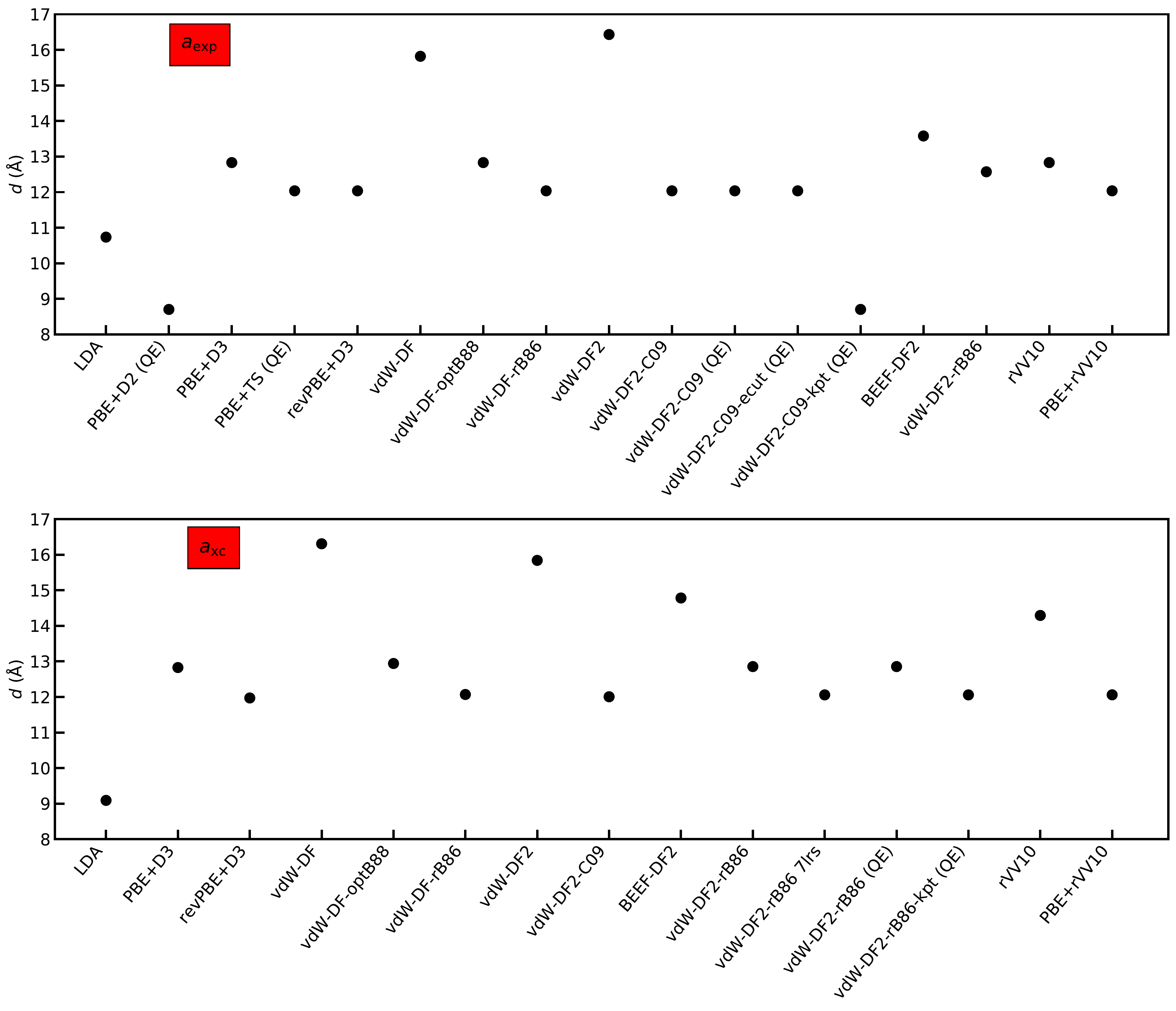}
\caption{\textbf{The apparent diameter of the depression $d$ in the gr/Ir(111) structures.} Evaluated using a geometric definition for the depression and assuming a spherical shape for it.}
\label{fig:gr_poresize}
\end{figure*}

To further characterise the geometries we evaluated an "apparent size of the depression". We define this using a geometric definition, where the depression is constituted from the B and N atoms -- or C in the case of graphene -- that are within the lower 20~\% of the extreme corrugation, or if $z_\mathrm{B/N} < z_\mathrm{min} + 0.2 (z_\mathrm{max} - z_\mathrm{min})$. We then define the apparent diameter of the depression e assuming as if the depression is circular and take its area as the area of the supercell multiplied by the ratio of the atoms in the depression by the total number of the atoms in the ad-layer. The results in Figures~\ref{fig:hBN_poresize} and \ref{fig:gr_poresize} indicate a variation in the apparent size of the depression. The diameter is very different in the two structures obtained with PBE+D2 and PBE+D3. In gr/Ir(111) the outliers are vdW-DF and vdW-DF2.

\section{Comparison of STM simulation and experiments}

\begin{figure*}[ht]
\includegraphics[width=\textwidth]{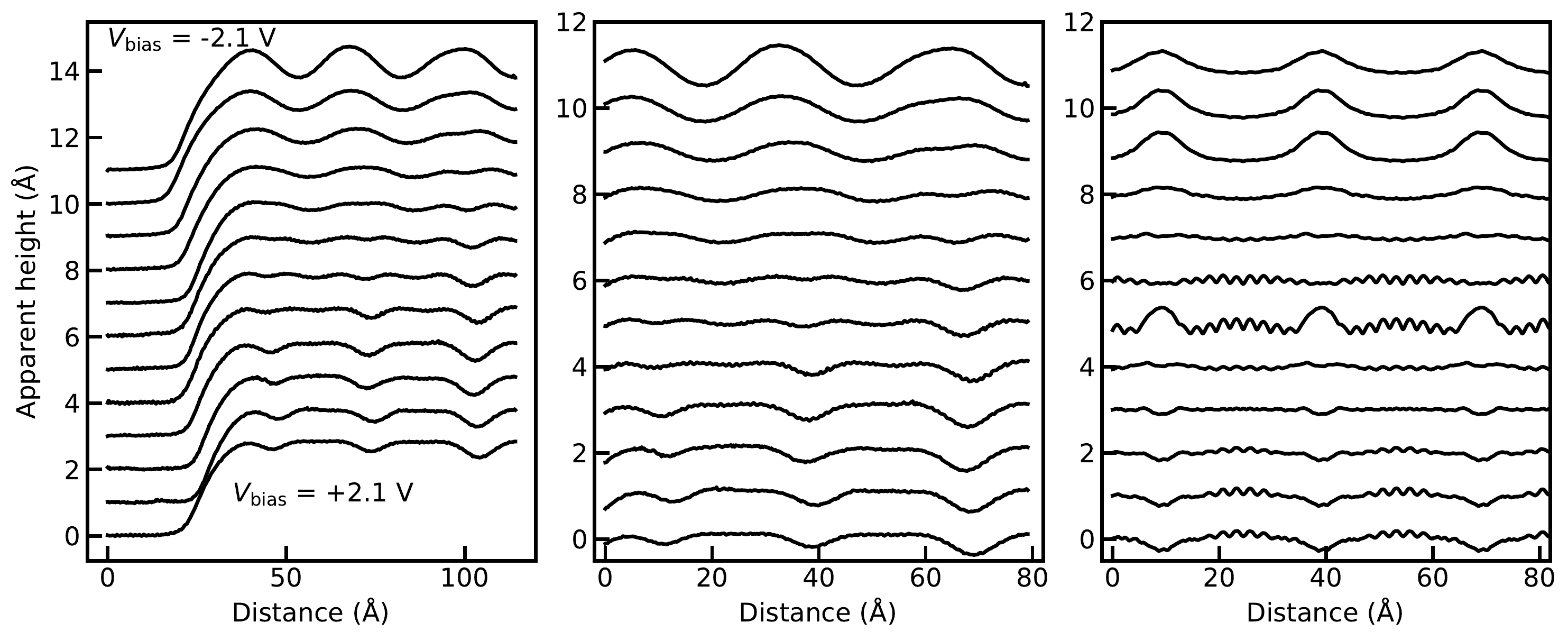}
\caption{\textbf{Comparison of experimental and calculated STM line profiles.} Experimental line profiles -- left and middle panel -- from
  Ref.~\citenum{Schulz_2014_a} and DFT-TH simulations -- right panel -- in
  the vdW-DF2-rB86 structure.}
\label{fig:hBN_STM}
\end{figure*}

We have earlier published bias-dependent STM experimental results and some DFT-derived simulated images within the $s$-type tip Tersoff-Hamann (TH) model.\cite{Tersoff_1998_a} In Figure~\ref{fig:hBN_STM}, we again show the experimental data and add the DFT-TH simulations at the same bias voltages $V_b$ from $\pm$0.1 to $\pm$2.1~V in steps of 0.4~V; the vdW-DF2-rB86 treatment of the XC term is applied here. The agreement is in general very good, only at $V_b=-0.1$~V there is probably too little local density of electrons for a reliable simulation of the STM-TH image and the corrugation is suddenly large due to numerical errors. At larger negative and intermediate positive values of
$V_b$, the shape and magnitude of the corrugation are in agreement, but at large positive $V_b$ the ``hills'' are narrower than in experiments, while the magnitude still reproduces the experimental trend. At the largest positive bias, $+2.1$~V, the lowest unoccupied electronic states of \textit{h}-BN start appearing in the DFT calculations\cite{Schulz_2014_a} because of the underestimation of the electronic band gap. Overall, the DFT-TH simulations reproduce the inversion of the corrugation between negative and positive $V_b$.

\section{Comparison of XPS simulation and experiments}

\begin{figure}[h!t!]
\centering
\includegraphics[width=\textwidth]{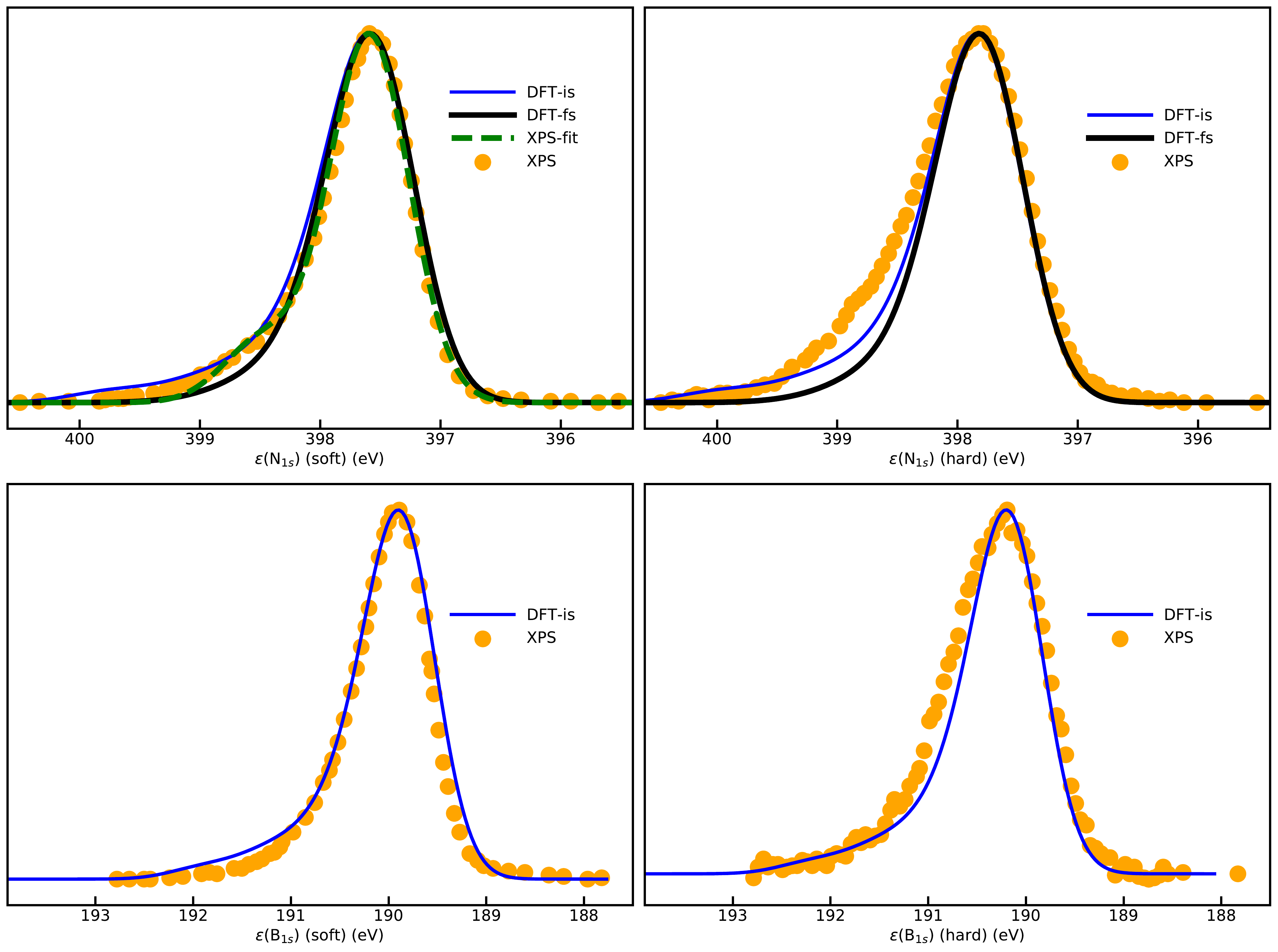}
\caption{\textbf{Comparison of experimental and simulated XPS spectra.} XPS experimental spectrum -- orange circles -- and a two-component -- dashed line -- fit from Ref.~\citenum{zumHagen_2016_a} together with simulations from DFT in the initial -- blue line -- and final state -- black line -- approximations.}
\label{fig:hBN_XPS}
\end{figure}

X-ray photoemission spectroscopy (XPS) experiments have also been conducted by Farwick zum Hagen et alia using both soft and hard X rays.\cite{zumHagen_2016_a} These yield information about the electronic structure at the vicinity of the atom from whose core level an electron is emitted. In the moir\'{e} structure a sum of all the sites of a given species is obtained; the experimental spectrum at the B$_{1s}$ and N$_{1s}$ levels together with a two-parametre fit by the Authors are shown in Figure~\ref{fig:hBN_XPS}, and compared with the DFT-QE results that were calculated using the vdW-DF-rB86 approximation. The latter were obtained either initial or final approximation. In the former the Kohn-Sham eigenvalues of the 1s orbital are subjected to Gaussian broadening, and the latter by removing one $1s$ electron from the core of each N atom at a time and averaging over the total energies of the different calculations, with a Gaussian broadening of the individual values; the energy scale of the DFT curve is shifted so that the peak is at the same position as in the experimental fit. The agreement is reasonably good, although in particular the wide asymmetry at the higher binding energies is different in the DFT results. This is similar as in the recent comparison of XPS and DFT in the system gr/Ru(0001).\cite{Silva_2018_JPCC_a} It remains to be investigated if the disagreement is due to the computational setup, or whether possibly the assumption of simple removal of a core-electron from the unit cell is not adequate.

\section{12-on-11 \textit{h}-BN/Ir(111) electronic structure}

\begin{figure}[ht]
\centering
\includegraphics[width=115mm]{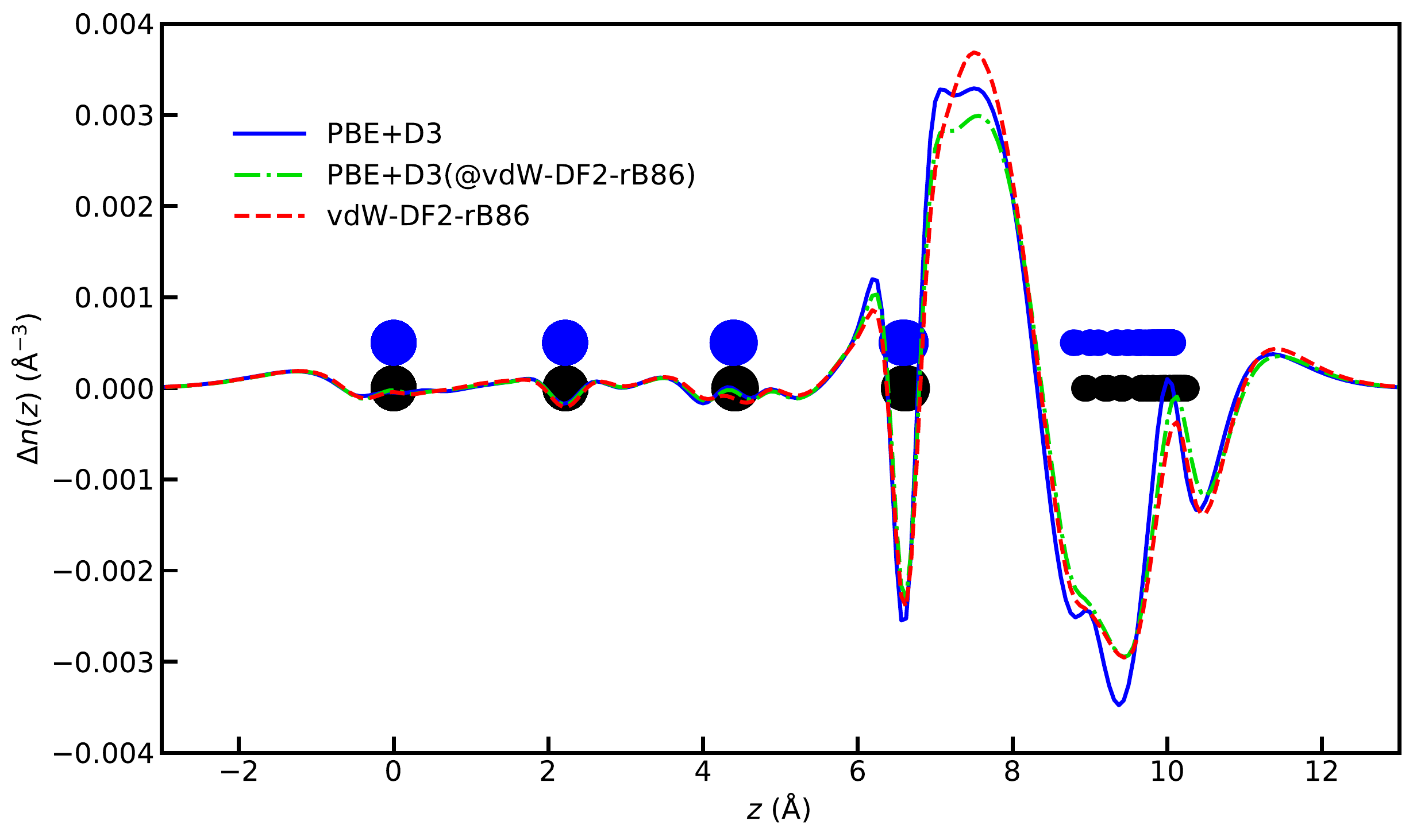}
\caption{\textbf{Difference in electron density $\Delta{}n(z)$ averaged parallel to the surface.} PBE+D3 and vdW-DF2-rB86 treatments of the XC were used, in PBE+D3(@vdW-DF2-rB86) the atomic positions were taken from the latter. The vertical positions of the atoms are shown with blue and black circles, respectively. Positive and negative values indicate enhancement and reduction of electron electron density upon the adsorption of \textit{h}-BN on the surface}
\label{fig:hBN_densdiff}
\end{figure}

In order to investigate the influence of different treatments of the XC to the electronic structure, we evaluated the difference in the electronic density due to the adsorption of the \textit{h}-BN onto the substrate; this density is then averaged along the surface: With the electron density $n_\mathrm{\textit{h}-BN/Ir}(\vec{r})$ of the full system, we subtract the individual electron densities $n_\mathrm{Ir}(\vec{r})$ and $n_\mathrm{\textit{h}-BN}(\vec{r})$, with the respective atomic coordinates of the full, adsorbed case,
$$
  \Delta{}n(z) = \int_{x,y} \left\{n_\mathrm{\textit{h}-BN/Ir}(\vec{r}) -
  \left[n_\mathrm{Ir}(\vec{r}) + n_\mathrm{\textit{h}-BN}(\vec{r})
    \right]\right\} \, dx\,dy \ .
$$
$\Delta{}n(z)$ evaluated with this formula from calculations with PBE+D3 and vdW-DF2-rB86 treatments of the XC are shown in Figure~\ref{fig:hBN_densdiff}. Also PBE+D3 electron density at the atomic positions of vdW-DF2-rB86 is shown (``PBE+D3(@vdW-DF2-rB86)''). The differences between the treatments are relatively small, even considering that in the case of PBE+D3 the electronic structure corresponds to the one from PBE, but with atomic positions forced away from the DFT minimum with the D3 term. A notable difference is the larger enhancement of the electron density in between the ad-layer and the substrate with vdW-DF2-rB86, in particular when the same atomic coordinates have been used.

%
%
\section{Results on \texorpdfstring{$(1\times{}1)$}{}-gr/Ir(111)}

\begin{figure}[h!t]
\centering
\includegraphics[width=.6\textwidth]{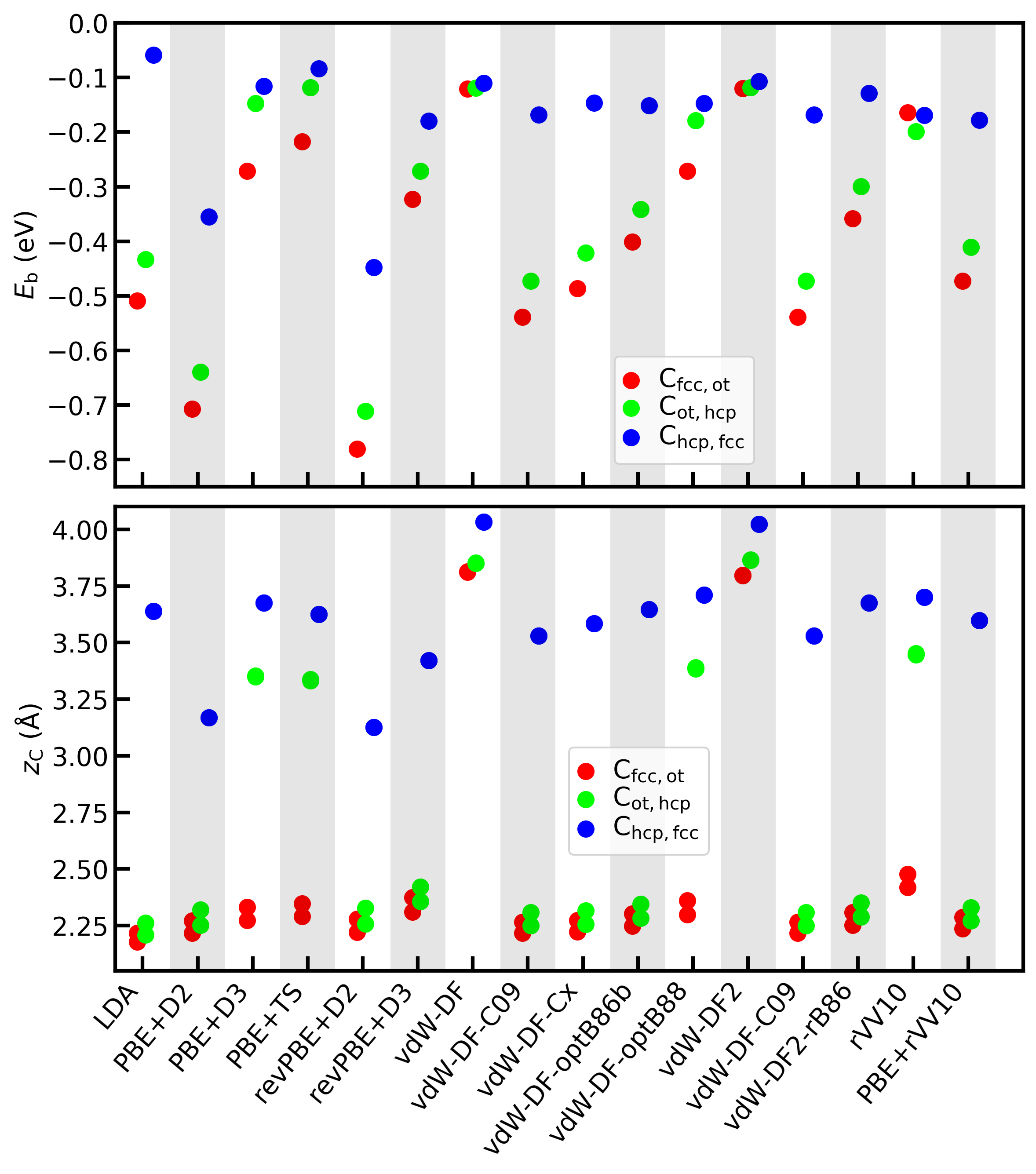}
\caption{\textbf{Results on commensurate $(1\times{}1)$ structure of gr/Ir(111).} (\textbf{A}) Binding energy $E_b$ and (\textbf{B}) height of $z_\mathrm{C}$ of C atoms above the top-most substrate layer in ($1\times{}1$) commensurate layer of graphene, at the experimental lattice constant of Ir(111) from QE-DFT calculations.}
\label{fig:hBN_1x1_gr_Ebind}
\end{figure}

We also calculated the forced-commensurate $(1\times{}1)$ structure of graphene adsorbed on Ir(111); our results of the binding energy and adsorption height and are shown in Figure \ref{fig:hBN_1x1_gr_Ebind} and Table
\ref{tab:gr_1x1_properties}; the experimental lattice constant has been used here. The strain in the graphene layer is even larger than in the \textit{h}-BN layer in the corresponding calculations, but we are here interested in qualitative effects. Indeed there are large differences between the different treatments of the XC effects, and again the original vdW-DF and vdW-DF2 give the weakest gr-substrate interaction, with negligible difference between the three studied lateral arrangements.

%
%

\bibliographystyle{apsrev4-1}
\bibliography{article-corrugation_pure}